# A Time Series Analysis-Based Stock Price Prediction Using Machine Learning and Deep Learning Models

## Sidra Mehtab[1] and Jaydip Sen[2]

[1]School of Computing and Analytics
NSHM Knowledge Campus, B. L. Saha Road
Kolkata 700104, West Bengal, INDIA
email: smehtab@acm.org

[2]Departmnet of Data Science and Artificial Intelligence
Praxis Business School, Bakrahat Road, Off Diamond Harbor Road
Rasapunja, Kolkata - 7000104, West Bengal, INDIA
email: jaydip.sen@acm.org

## Abstract

Prediction of future movement of stock prices has always been a challenging task for researchers. While the advocates of the *efficient market hypothesis* (EMH) believe that it is impossible to design any predictive framework that can accurately predict the movement of stock prices, there are seminal work in the literature that have demonstrated that the seemingly random movement patterns in the time series of a stock price can be predicted with a high level of accuracy. The design of such predictive models requires the choice of appropriate variables, right transformation methods of the variables, and tuning of the parameters of the models. In this paper, we present a very robust and accurate framework of stock price prediction that consists of an agglomeration of statistical, machine learning, and deep learning models. We use daily stock price data, collected at five minutes intervals of time, of a very well-known company that is listed in the National Stock Exchange (NSE) of India. The granular data is aggregated into three slots in a day, and the aggregated data is used for training and building the forecasting models. We contend that the agglomerative approach of model building that uses a combination of statistical, machine learning, and deep learning approaches, can very effectively learn from the volatile and random movement patterns in stock price data. This effective learning will lead to the building of very robust training of the models that can be deployed for short-term forecasting of stock prices, and prediction of stock movement patterns. We build eight classification and eight regression models based on statistical and machine learning approaches. In addition to these models, two deep learning-based regression models using a long-and-short-term memory (LSTM) network and a convolutional neural network (CNN) have also been built. Extensive results have been presented on the performance of these models, and results are critically analyzed.

## 1. Introduction

Prediction of future movement patterns of stock prices has been a widely researched area in the literature. While there are proponents of the efficient market hypothesis who believe that it is impossible to predict stock prices, there are also propositions that demonstrated that if correctly formulated and modeled, the prediction of stock prices can be done with a fairly high level of accuracy. The latter school of thought focused on the construction of robust statistical, econometric, and machine learning models based on the careful choice of variables and appropriate functional forms or models of forecasting. There are propositions in the literature that are based on time series analysis and decomposition for forecasting future values of stocks. In this regard, several propositions have been presented in the literature for stock price forecasting following a time series decomposition approach. (Sen & Datta Chaudhuri, 2018a; Sen, 2018b; Sen, 2018c; Sen & Datta Chaudhuri, 2016a; Sen & Datta Chaudhuri, 2016b). There is also an extent of literature that deals with various technical analysis of stock price movements. Propositions also exist for mining sock price patterns using various important indicators like *Bollinger Bands*, *moving average convergence divergence* (MACD), *relative strength index* (RSI), *moving average* (MA), *stochastic momentum index* (SMI), etc. There are also well-known patterns like *head and shoulders pattern*, *inverse head and shoulders pattern*, *triangle*, *flag*, *Fibonacci fan*, *Andrew's Pitchfork*, etc., which are exploited by traders for investing intelligently in the stock market. These approaches provide the user with visual manifestations of the indicators which help the ordinary investors to understand which way stock prices are more likely to move soon. In this thesis, we propose a granular approach to forecasting stock price and the price movement pattern by combining several statistical, machine learning, and deep learning methods of prediction on technical analysis of stock prices. We present several approaches for short-term stock price movement forecasting using various classification and regression techniques and compare their performance in the prediction of stock price movement and stock price values. We believe this approach will provide several useful information to the investors in the stock market who are particularly interested in short-term investments for profit. This work is a modified and extended version of our previous work (Mehtab & Sen, 2019). In the present work, we have presented a predictive framework that aggregates eight classification and eight regression models including a *long-and short-term memory* (LSTM)-based advanced deep learning model, and four variants of convolutional neural network (CNN)-based forecasting models.

The objective of our work is to take stock price data at five minutes intervals from the National Stock Exchange (NSE) of India and develop a robust forecasting framework for the stock price movement. We contend that such a granular approach can model the inherent dynamics and can be fine-tuned for immediate forecasting of stock price or stock price movement. Here, we are not addressing the problem of forecasting of long-term movement of the stock price. Rather, our framework will be more relevant to a trade-oriented framework.

The rest of the paper is organized as follows. Section 2 presents a comprehensive review of the literature on stock price movement modeling and prediction. In Section 3, we present a detailed discussion on the methodology that we have followed in this work. Section 4 provides a brief discussion on the working principles of the classification and the regression models in machine learning that we have used in this work. In Section 5, we present a summary of two deep learning-based models – LSTM-based deep learning model for regression and four variants of CNN-based forecasting models - that we have also used in our predictive model. Section 6 presents a detailed discussion on the performance of machine learning and deep learning models. A comparative analysis of the performances of the models is also presented in this section. Finally, Section 7 concludes the paper.

## 2. Related Work

The literature attempting to prove or disprove the efficient market hypothesis can be classified into three strands, according to the choice of variables and techniques of estimation and forecasting. The first strand consists of studies using simple regression techniques on cross-sectional data (Enke et al., 2011; Ma & Liu,

2008; Khan et al., 2018; Ivanovski, 2016; Sen & Datta Chaudhuri, T, 2016c. The second strand of the literature has used time series models and techniques to forecast stock returns following economic tools like *autoregressive integrated moving average* (ARIMA), Granger causality test, *autoregressive distributed lag* (ARDL), and *quantile regression* (QR) to forecast stock prices (Ariyo et al., 2014; Jammalamadaka et al., 2019; Jarrett & Kyper, 2011; Mondal et al., 2014; Sen & Datta Chaudhuri, 2017; Xiao et al., 2014). The third strand includes work using machine learning, deep learning, and natural language processing for the prediction of stock returns (Mostafa, 2010; Dutta et al., 2006; Mehtab & Sen, 2019; Mehtab & Sen, 2020a; Mehtab & Sen, 2020b; Mehtab & Sen, 2020c; Mehtab et al., 2020d; Mehtab et al., 2020e; Mehtab & Sen, 2021; Porshnev et al., 2013; Obthong et al., 2020; Sen, 2018d; Tang & Chen, 2018; Wang et al., 2018; Zhou & Fan, 2019; Wu et al., 2008).

Among some of the recent propositions in the literature on stock price prediction, Mehtab and Sen have demonstrated how machine learning and *long- and short-term memory* (LSTM)-based deep learning networks can be used for accurately forecasting NIFTY 50 stock price movements in the National Stock Exchange (NSE) of India (Mehtab & Sen, 2019). The authors used the daily stock prices for three years from January 2015 till December 2017 for building the predictive models. The forecast accuracies of the models were then evaluated based on their ability to predict the movement patterns of the close value of the NIFTY index on a time horizon of one week. For testing, the authors used NIFTY 50 index values for January 2018 till June 2019. To further improve the predictive power of the models, the authors incorporated a sentiment analysis module for analyzing the public sentiments on Twitter on NIFTY 50 stocks. The output of the sentiment analysis module is fed into the predictive model in addition to the past NIFTY 50 index values for building a very robust and accurate forecasting model. The sentiment analysis module uses a *self-organizing fuzzy neural network* (SOFNN) for handling non-linearity in a multivariate predictive environment.

Mehtab and Sen recently proposed another approach to stock price and movement prediction using *convolutional neural networks* (CNN) on a multivariate time series (Mehtab & Sen, 2020). The predictive model proposed by the authors exploits the learning ability of a CNN with a *walk-forward validation* ability to realize a high level of accuracy in forecasting the future NIFTY index values, and their movement patterns. Three different architectures of CNN are proposed by the authors that differ in the number of variables used in forecasting, the number of sub-models used in the overall system, and the size of the input data for training the models. The experimental results indicated that the CNN-based multivariate forecasting model was highly accurate in predicting the movement of NIFTY index values with a weekly forecast horizon. Use of LSTM networks in stock price prediction has also been proposed (Sen et al., 2021a; Sen et al., 2021b).

Zhang et al. proposed the application of a multilayer backpropagation (BP) neural network in financial data mining (Zhang et al., 2007). The proposed scheme was a modified neural network-based forecasting model that carries out intelligent mining tasks. The system was capable of making robust forecasting on the buying and selling signs according to the prediction of future trends in the stock market. The simulation results on seven years of data of the Shanghai composite index indicated that the return achieved by the system is about three times that achieved by the buy-and-hold strategy.

Basalto et al. proposed an approach based on a pair-wise clustering to analyze the Dow Jones Index companies to identify similar temporal behavior of the traded stock prices (Basalto et al., 2005). The main goal of the authors was to investigate and understand the dynamics that govern companies' stock prices. The proposed scheme deployed a pairwise version of the chaotic map algorithm that was executed based on correlation coefficients between the financial time series to find similarity measures for clustering the temporal patterns. The resultant dynamics of such systems formed the clusters of companies that belong to different industrial branches. These clusters of companies can be gainfully exploited to optimize portfolio construction.

Chen et al. have proposed an approach for constructing a model for predicting the direction of return on the Taiwan Stock Exchange Index (Chen et al., 2003). The authors contended that the stock trading guided by robust forecasting models were more effective and usually led to a higher return on investment. To construct a robust forecasting model, the authors built and trained a probabilistic neural network (PNN) using historical stock market data. The forecasted output of the model was applied to form various index trading strategies, and the effectiveness of those strategies was compared with those generated by the buy and hold strategy, the investment strategies formed using the output of a random walk model, and the parametric generalized method of moments (GMM) with a Kalman filter. The results showed that the investment strategies made using the output of the PPN yielded the highest return of investment in the long-run.

de Faria et al. illustrated a predictive model using a neural network and an adaptive exponential smoothing (AES) method for forecasting the movements of the principal index of the Brazilian stock market (de Faria et al., 2009). The authors compared the forecasting performance of both the neural network and the exponential smoothing models with a particular focus on the sign of the market returns. While the simulation results showed that both methods were equally efficient in predicting the index returns, the neural network model was found to be more accurate in predicting the market movement than the adaptive exponential smoothing method.

Leigh et al. proposed the use of linear regression and simple neural network models for forecasting the stock market indices in the New York Stock Exchange during the period 1981-1999 (Leigh et al., 2005). The proposed scheme by the authors used a template matching mechanism based on statistical pattern recognition that efficiently and accurately identified spikes in the trading volumes. A threshold limit for the spike in volume was identified, and the days on which the traded volume exhibited significant spikes were identified. A linear regression model was applied to forecast the future change in price based on the historical price, traded volume, and the prime interest rate.

Shen et al. proposed a novel scheme that was based on a tapped delay neural network (TDNN) with an ability of adaptive learning and pruning for forecasting on a non-linear time series of stock price values (Shen et al., 2007). The TDDN model was trained by a recursive least square (RLS) technique that involved a tunable learning-rate parameter that enables faster network convergence. The trained neural network model was optimized using a pruning algorithm that reduced the possibility of overfitting of the model. The experimental results in a simulated environment clearly showed that the pruned model had a reduced complexity, faster execution, and improved prediction accuracy.

Ning et al. proposed a scheme of stock index prediction that was based on a chaotic neural network (Ning et al., 2009). Data from a Chinese stock market and a Shenzhen stock market were used for building the model. The non-linear, stochastic, and chaotic patterns in the stock market indices were learned by the chaotic neural network, and the learnings of the chaotic neural network were gainfully applied in forecasting future index values of the stock markets.

Hanias et al. conducted a study to predict the daily stock exchange price index of the Athens Stock Exchange (ASE) using a neural network with backpropagation (Hanias et al., 2012). The neural network was used to make a multistep forecasting for nine days and yielded a very low mean square error (MSE) value of 0.0024.

Wu et al. proposed an ensemble model of prediction using support vector machines (SVM) and artificial neural networks (ANN) for predicting stock prices (Wu et al., 2008). The forecasting performance of the ensemble model was compared with those of the SVM model and the ANN model. It was observed by the authors that the ensemble approach produced more accurate results than the other two models.

Liao et al. carried out a study on the stock market investment issues on the Taiwan stock market (Liao et al., 2008). The scheme involved two phases. In the first phase, the apriori algorithm was used to identify the association rules and knowledge patterns about stock category association and possible stock category investment collections. After the association rules were successfully mined, in the second phase, the k-means clustering algorithm was used to identify the various clusters of stocks based on their association patterns. The authors also proposed several possible stock market portfolio alternatives under various clusters of stocks.

Zhu et al. hypothesized that there is a significant bidirectional nonlinear causality between stock returns and trading volumes (Zhu et al., 2008). The authors proposed the use of a neural network-based scheme for forecasting stock index movements. The model was further enriched by the inclusion of different combinations of indices and component stocks' trading volumes as inputs. NASDAQ, DJIA, and STI data of stock prices and volume of transactions were used in training the neural network. The experimental results demonstrated that the augmented neural networks with trading volumes lead to improvements in forecasting performance under different terms of the forecasting horizon.

Bentes et al. presented a study on the long memory and volatility clustering for the S&P 500, NASDAQ 100, and Stoxx 50 indexes to compare the US and European markets (Bentes et al., 2008). The authors compared the performance of two different approaches. The first approach was based on the traditional approaches using generalized autoregressive conditional heteroscedasticity GARCH(1, 1), IGARCH(1, 1), and FIGARCH (1, d, 1), while the second approach exploited the concept of entropy in the Econophysics. In the second approach, three different measures were considered by the authors in the study. The three measures were Shannon, Renyi, and Tsallis measures. The results obtained using both the approaches elicited the existence of nonlinearity and volatility of SP 500, NASDAQ 100, and Stoxx 50 indexes.

Chen et al. demonstrated how the random and chaotic behavior of stock price movements can be very effectively modeled using a local linear wavelet neural network (LLWNN) technique (Chen et al., 2005). The proposed wavelet-based model was further optimized using a novel algorithm, which the authors referred to as estimation of distribution algorithm (EDA). The purpose of the model was to accurately predict the share price for the following trade day given the opening, closing, and maximum values of the stock price for a particular day. The study revealed an interesting observation - even for a time series that exhibited an extremely high level of random fluctuations in its values, the model could extract some very important features from the opening, closing, and the maximum values of the stock index that enabled an accurate prediction of its future behavior.

Dutta et al. illustrated how ANN models could be applied in forecasting the Bombay Stock Exchange's SENSEX weekly closing values from January 2002 to December 2003 (Dutta et al., 2006). The proposed approach by the author involved building two neural networks each consisting of three hidden layers, in addition to the input and the output layers. The input values to the first neural network were: (i) the weekly closing values, (ii) the 52-week moving average of the weekly closing SENSEX values, (iii) the 5-week moving average of the closing values, and (iv) the 10-week oscillator values for the past 200 weeks. On the other hand, the second network was provided with the following input values: (i) weekly closing value of SENSEX, (ii) the moving average of the weekly closing values computed on the 52-week historical data, (iii) the moving average of the closing values computed on the 5-week historical data, and (iv) the volatility of the SENSEX records computed on 5-week basis over the past 200 weeks. The forecasting performance of the two neural networks was compared using their root mean square error (RMSE) and mean absolute error (MSE) values on the test data. To test the networks, the weekly closing SENSEX values for the period of January 2002 to December 2003 were used.

Hammad et al. demonstrated that an artificial neural network (ANN) model can be trained to converge to an optimal solution while it maintains a very high level of precision in the forecasting of stock prices (Hammad et al., 2009). The proposed scheme was based on a multi-layer feedforward neural network model that used the back-propagation algorithm. The model was used for forecasting the Jordanian stock prices. The authors demonstrated simulations using MATLAB that were carried on seven Jordanian companies from the service and manufacturing sectors. The accuracy of the model in forecasting stock price movement was found to be very high.

Tseng et al. utilized various approaches including the traditional time series decomposition (TSD) model, HoltWinters (H/W) exponential smoothing with trend and seasonality models, Box-Jenkins (B/J) models using autocorrelation and partial autocorrelation, and neural network-based models (Tseng et al, 2012). The authors trained the models on the stock price data of 50 randomly chosen stocks during the period: September 1, 1998 - December 31, 2010. To train the models, 3105 observations based on closed prices of the stocks were used. The testing of the model was carried out on data spanning over 60 trading days. The study showed that the forecasting accuracies were higher for B/J, H/W, and normalized neural network models. The errors associated with the time series decomposition-based model and the non-normalized neural network models were found to be higher.

Senol and Ozturan illustrated that ANN can be used to predict stock prices and their direction of changes (Senol & Ozturan, 2008). The result was promising with a forecast accuracy of 81% on average.

Fu et al. presented an approach that represented the data points in a financial time series according to their importance (Fu et al., 2008). Using the ranked data points based on their importance, a tree was constructed that enabled incremental updating of data in the time series. The scheme facilitated the representation of a large-sized time series in different levels of details, and also enabled multi-resolution dimensionality reduction. The authors have presented several evaluation methods of data point importance, a novel method of updating a time series, and two-dimensionality reduction approaches. Extensive experimental results are also presented demonstrating the effectiveness of all propositions.

Phua et al. presented a predictive model using neural networks with genetic algorithms for forecasting stock price movements in the Singapore Stock Exchange (Phua et al., 2001). The forecasting accuracy of the predictive model was found to be 81% on the test dataset indicating that the model was moderately effective in its forecasting job.

Moshiri and Cameron described a back propagation-based neural network and a set of econometric models to forecast inflation levels (Moshiri, & Cameron, 2010). The set of econometric models proposed by the authors included the following: (i) Box-Jenkins autoregressive integrated moving average (ARIMA) model, (ii) vector autoregression (VAR) model, and (iii) Bayesian vector autoregression (BVAR) model. The forecasting accuracies of the three models were compared with the hybrid back propagation network (BPN) model proposed by the authors. To test the models, three different values of the forecasting horizon were used: one month, two months, and twelve months. With the root mean square error (RMSE) and the mean absolute error (MAE) as the two metrics, the authors observed that the performance of the hybrid BPN was superior to the other econometric models.

Asghar et al. contend that the stock price data acquired from social media and the websites storing financial data, is usually sparse, and often the predictors used for building the stock price prediction models are poorly chosen (Asghar et al., 2019). This leads to poor performance of the models. The authors propose a multiple regression model that systematically selects the predictors, and hence yields a high accuracy in forecasting. The system also has a user-friendly interface that provides a rich user experience. However, the model is too simple to effectively handle time series containing highly volatile stock prices.

Park et al. present a model for stock price prediction that is based on multiple regression (Park et al., 2010). The model has the capability of estimating the risk associated with a stock price. However, the framework is too simple and the assumptions made in designing the model may not be holding good in real-world times series of volatile stock price movements.

Yan et al. propose a hybrid predictive model that consists of a multiple linear regression model and a backpropagation (BP) neural network model for predicting the movements of the stock prices (Yan et al., 2019). The results make it evident that the BP neural network is more accurate than its multiple regression counterpart. However, the effectiveness of the BP neural network model on a highly granular and volatile stock price data is questionable.

Vantuch et al. propose an ARIMA-based stock price forecasting model (Vantuch & Zelinka, 2014). The ARIMA model is further fine-tuned using evolutionary algorithms that are based on *genetic algorithms* (GA) and *particle swarm optimization* (PSO). The model is found to have a very high level of accuracy in forecasting. However, its applicability on a very highly granular time series data with a very short forecast horizon is not addressed by the authors.

Ning et al. investigate the relationship between several macroeconomic variables, e.g., interest rate, money supply, exchange rate, inflation rate, etc., and their effect on stock returns in Hong Kong and Shanghai (Ning et al., 2019). The relationships are tested using *arbitrage pricing theory* (APT), *vector error correction model* (VECM), and the Granger causality test. The results elicit an important observation – investors should have long-term investments in the Chinese stock market for getting a good return on their investments, while for the Hong Kong stock markets, the case is just the opposite.

Bao et al. present a hybrid deep learning framework for stock price prediction that consists of three components: (i) wavelet transform (WT), (ii) stacked autoencoders (SAEs), and (iii) long-and short-term memory (LSTM) gates (Bao et al., 2017). Initially, the time series of the stock price data is decomposed the WT for denoising of the data. The denoised data is passed on to the SAEs that extract deep features from the data, which are, then, passed into the LSTM module for predicting the future stock prices. The model is found to yield very high accuracy. However, the model is evaluated on a stock price data that has a daily frequency. Hence, it is not suitable for intra-day investment decisions.

Vargas et al. propose a sophisticated deep learning model for the detection and analysis of complex patterns and interactions among stock price data (Vargas et al., 2017). The deep learning framework consists of a CNN and a RNN model. The results show that CNN is more effective in catching the semantic meaning from the text inputs to the model, while RNN is superior in understanding the context information and modeling the temporal characteristics for the stock price prediction.

The major drawback of the existing propositions in literature for stock price prediction is their inability to predict stock price movement in a short-term interval. The current work attempts to address this shortcoming by exploiting the learning ability of a gamut of machine learning and two deep neural networks in stock price movement modeling and prediction.

## 3. Methodology

In Section 1, we mentioned that the goal of this work is to develop a robust forecasting framework for the short-term price movement of stocks. We use the Metastock tool for collecting data on the short-term price movement of stocks (Metastock). Particularly, we collected the stock data for the company – *Godrej Consumer Products Ltd*. The data is collected at every 5 minutes interval in a day, for all the days in which the National Stock Exchange (NSE) was operational during the years 2013 and 2014. The raw data for each stock consisted of the following variables: (i) *date*, (ii) *time*, (iii) *open* value of the stock, (iv) *high* value of the stock, (v) *low* value of the stock, (vi) *close* value of the stock, and (vii) *volume* of the stock traded in a given interval. The variable *time* refers to the time instance at which the stock values are noted as each record is collected at 5 minutes interval of time. Hence, the time interval between two successive records in the raw data was 5 minutes. The raw data in this format is collected for the stock Godrej Consumer Products. for two years. In addition to the seven variables in the raw data that we have mentioned above, we also collected the NIFTY index at 5 minutes interval for the same period of two years, To capture the overall market sentiment at each time instant, so that more accurate and robust forecasting can be made using the combined information of historical stock prices and the market sentiment index. Therefore, the raw data for both the stocks now consists of seven variables. As 5 minutes interval is too granular, we make some aggregation of the raw data. We break the total time interval in a day into three slots as follows: (1) *morning slot* that covers the time interval 9:00 AM till 11:30 AM, (2) *afternoon slot* that covers the time interval 11:35 AM till 1:30 PM, and (3) *evening slot* that covers the time interval 1:35 PM till the time of closure of NSE in a given day. Hence, the daily stock information now consists of three records, each record containing stock price information for a time slot.

Using the eight variables in the raw data, and incorporating the aggregation of data using the time slots, we create eleven derived variables and compute their values. These derived variables are used as the input variables for building the predictive models for forecasting the stock price and the stock movement. We followed two approaches to stock price forecasting - *regression* and *classification*. The difference in these two approaches lied in the way the response variable *open_perc* was used in the model building process. This point will be described in detail later in this Section.

The following are the eleven variables that are computed for designing the machine learning models.

*month*: This is a numeric variable that refers to the month for a given stock price record. The twelve months are assigned numeric codes of 1 through 12, with January being coded as 1, and the month of December assigned with a code of 12.

*day_month*: This numeric variable denotes the particular day of a given month to which a stock price record corresponds. The value of this variable lies in the interval [1, 31]. For instance, if the date for a stock price record is 22nd May 2013 then the *day_month* variable for that record will be assigned a value of 22.

*day_week*: This is a numeric variable that corresponds to the day of the week for a given stock price record. The five days in a week on which the stock market remain open are assigned numeric codes of 1 through 5, with Monday being coded as 1, while Friday is assigned a code of 5.

*time*: This numeric variable refers to the time slot to which a stock price record belongs. There are three-time slots in a day - *morning*, *afternoon,* and *evening*. The slots are assigned codes 1, 2, and 3 respectively. For example, if a stock price record refers to the time point 3:45 PM, the variable *time* will be assigned a value of 3 for the stock price record.

*open_perc*: it is a numeric variable that is computed as a percentage change in the value of the *open* value of the stock over two successive time slots. The computation of the variable is done as follows. Suppose, we have two successive slots: $S_1$ and $S_2$. Both of them consist of several records at five minutes intervals of time. Let the open price of the stock for the first record of $S_1$ is $X_1$ and that for $S_2$ is $X_2$. The *open_perc* for the slot $S_2$ is computed as $(X_2 - X_1)/X_2$ in terms of percentage.

*high_perc*: it is a numeric value that is computed as the difference between the *high* values of two successive slots. The computation is identical to that of *open_perc* except for the fact that *high* values are used in this case instead of the *open* values.

*low_perc*: it is a numeric value that is computed as the difference between the *low* values of two successive slots. For two successive slots $S_1$ and $S_2$, first we compute the mean of all *low* values of the records in both the slots. If $L_1$ and $L_2$ refer to the mean of the *low* values for $S_1$ and $S_2$ respectively, then *low_perc* for $S_2$ is computed as $(L_2 - L_1)/L_2$ in terms of percentage.

*close_perc*: it is a numeric value that is computed as the difference between the *close* values of two successive slots. Its computation is similar to the *open_perc* variable, except for the fact that we use the *close* values in the slots instead of the *open* values.

*vol_perc*: it is a numeric value that is computed as the difference between the *volume* values of two successive slots. For two successive slots $S_1$ and $S_2$, we compute the mean values of *volume* for both the slots, say $V_1$ and $V_2$ respectively. Now, the *vol_perc* for $S_2$ is computed as $(V_2 - V_1)/V_2$ in terms of percentage.

*nifty_perc*: it is a numeric variable that is computed as a percentage change in the NIFTY index over two successive time slots. The computation of the variable is done as follows. We compute the means of the NIFTY index values for two successive time slots $S_1$ and $S_2$. Let us assume the means are $M_1$ and $M_2$ respectively. Then the *nifty_perc* for the slot $S_2$ is computed as $(M_2 - M_1)/M_2$ in terms of percentage.

*range_diff*: The value of this numeric variable is obtained by computing the difference in the *range* values of two consecutive time slots. The range value for a given slot is the difference between its *high* and *low* values. If $S_1$ and $S_2$, denote two consecutive slots, and if $H_1$, $H_2$, $L_1$, and $L_2$ respectively represent the *high* and the *low* values of the slots $S_1$ and $S_2$, then the range value for $S_1$ is $R_1 = (H_1 - L_1)$ and for $S_2$ is $R_2 = (H_2 - L_2)$. The *range_diff* for the slot $S_2$ is computed as $(R_2 - R_1)$.

After we compute the values of the above eleven variables for each slot for both the stocks for the time frame of two years (i.e., 2013 and 2014), we develop the forecasting framework. As mentioned earlier, we followed two broad approaches in the forecasting of the stock prices - *regression* and *classification*.

In the *regression* approach, based on the historical movement of the stock prices we predict the stock price in the next slot. We use *open_perc* as the response variable, which is a continuous numeric variable. The objective of the regression technique is to predict the *open_perc* value of the next slot, given the stock movement pattern and the values of the predictors till the previous slot. In other words, if the current time slot is $S_1$, the regression techniques will attempt to predict *open_perc* for the next slot $S_2$. If the predicted *open_perc* is positive, then it will indicate that there is an expected rise in the stock price in $S_2$, while a negative *open_perc* will indicate a fall in the stock price in the next slot. Based on the predicted values, a potential investor can make his/her investment decision in stocks.

In the *classification* approach, the response variable *open_perc* is a discrete variable belonging to one of the two classes – "0" or "1". For designing the classification-based forecasting models, we converted *open_perc* into a categorical variable that assumes one of the two values "0" and "1". While "0" indicates a negative or a zero *open_perc* value, "1" indicates a positive *open_perc* value. Hence, if the current slot is $S_1$ and if the predictive model expects a rise in the *open_perc* value in the next slot $S_2$, then the *open_perc* value for $S_2$ will be "1". An expected negative value of the *open_perc* in the next slot will be indicated by a "0" value for the response variable.

For both classification and regression approaches, we studied three cases. In the following, these scenarios are described in detail.

*Case I*: We used the data for the year 2013 which consisted of 19, 385 records at five minutes intervals. These records were aggregated into 745-time slot records for building the predictive model. We used the same dataset for testing the forecast accuracy of the models for the stock of Godrej Consumer Products Ltd. and carried out a comparative analysis of all the models.

*Case II*: We used the data for the year 2014 which consisted of 18, 972 records at five minutes intervals. These granular data were aggregated into a 725-time slot record for building the predictive model. We used the same dataset for testing the forecast accuracy of the models and carried out an analysis of the performance of the predictive models.

*Case III*: We used that data for 2013 as the training dataset for building the models and test the models using the data for the year 2014 as the test dataset. We, again, carried out an analysis of the performance of different models in this approach.

We build eight classification models and ten regression models for developing our forecasting framework. The classification models are: (i) *logistic regression*, (ii) *k-nearest neighbor* (iii) *decision tree*, (iv) *bagging*, (v) *boosting*, (vi) *random forest*, (vii) *artificial neural network*, and (viii) *support vector machines*. For measuring accuracy and effectiveness in these approaches, we use several metrics such as *sensitivity*, *specificity*, *positive predictive value*, *negative predictive value*, *classification accuracy*, and *F1 score*. *Sensitivity* and *positive predictive value* are also known as *recall* and *precision* respectively.

The ten regression methods that we build are: (i) *multivariate regression*, (ii) *multivariate adaptive regression spline* (MARS), (iii) *decision tree*, (iv) *bagging*, (v) *boosting*, (vi) *random forest*, (vii) *artificial neural network*, (viii) *support vector machine*, (ix) *long- and short-term memory* (LSTM) network, (x) *convolutional neural network* (CNN).

While all the classification techniques are machine learning-based approaches, two regression techniques, i.e., *long- and short-term memory* (LSTM) network, and *convolutional neural network* (CNN) – based approaches are deep learning methods. For comparing the performance of the regression methods, we use three metrics: (i) *root mean square error* (RMSE), (ii) *correlation coefficient* between the actual and predicted values of the response variable, e.g., *open_perc*, and (iii) *the number of cases in which the predicted and the actual values of open_perc differed in their signs*.

## 4. Machine Learning Models

The eight classification models that we built are discussed in detail in this Section.

*Logistic Regression*: Logistic regression is a classification technique, and hence in building this model, we transform the response variable *open_perc* into a discrete (i.e., categorical) variable that can assume two discrete values - "0" or "1". We convert all negative or zero values of *open_perc* to the class "0", and all positive values to class "1". We use the function *glm* in R for building the logistic regression model with three parameters passed into the function: (i) the first parameter is the formula which is: *open_perc ~.* to include *open_perc* as the response variable and all the remaining variables as the predictors, (ii) the second parameter is: *family = binomial* – this indicates that model is a binary logistic regression involving two classes, and (iii) the third parameter is the R data object containing the *training data set*. We use the *predict* function in R to compute the probabilities of the test records belonging to the two classes. A threshold probability value of 0.5 is assumed. In other words, when the probability of a record belonging to a class exceeds 0.5, the record is classified into that particular class.

*K-Nearest Neighbor*: The KNN model follows what is known as *instance-based learning*. Based on the training, the classification for a new unclassified record is found simply by comparing it to the most similar records in the training set. The value of *k* determines how many closest similar records in the training data set is considered for classifying a test data set record. We use the function *knn* defined in the R library *class* to carryout KNN classification in the stock price data. The data is normalized using *min-max normalization* before applying the *knn* function so that all predictors are *scaled down* into the same range of values. Different values of *k* are tried out for building the models, and the value of *k* = 3 is finally chosen. This value of *k* is found to yield the best performance with the minimum possibility of *model overfitting*.

*Decision Tree*: The *classification and regression tree* (CART) algorithm produces decision trees that are strictly binary so that there are exactly two branches at each node. The algorithm recursively partitions the records in the training data set into subsets of records with similar values for the target attributes. The trees are constructed by carrying out an exhaustive search on each node for all available variables and all possible splitting values and selects the optimal split based on some goodness of split criteria. We use the *tree* function defined in the *tree* library of R for the classification of the stock records.

*Bagging*: *Bootstrap Aggregation* (Bagging) is an ensemble technique. It works as follows: Given a set *D*, of *d* tuples, for iteration *i*, a training set $D_i$ of *d* tuples is sampled *with replacement* from the original set of *D* tuples. Each training set represents a *bootstrap sample*. Since the samples are *simple random samples with replacement*, some records (i.e., tuples) in *D* may not get a chance to be included in $D_i$, while some tuples may get included in more than one sample. A classifier model $M_i$ is trained on the information contained in each training set $D_i$. For classifying an unknown tuple *X* in the *out-of-sample set* (i.e., in the test dataset), each classifier $M_i$ is asked to return its class predictions. The classification result of each of the trained classifier is considered as *one vote*. The bagging classifier counts the votes and finally assigns the class with the *maximum number of votes* to the tuple *X*. For carrying out classification on stock price data, we use *bagging* function defined in the *ipred* library of R. The value of the parameter *nbag* is taken as 25. This specifies the number of bagging classifiers used in simple majority voting.

*Boosting*: Unlike *bagging*, *boosting* assigns weights to each tuple in a training dataset. Based on the training dataset, *k* classification models are built iteratively. However, all the classifiers are not given equal importance in the final classification decision. Unlike bagging which uses *simple majority voting* among the classifiers, boosting uses a *weighted majority voting mechanism*. After a classifier $M_i$ is constructed, the weights assigned to the classifiers are updated before building the subsequent classifier $M_{i+1}$. After the completion of the current iteration, the classifiers that could correctly classify the tuples which were misclassified in the previous round are assigned higher weights before the next iteration of classifier construction starts. After the completion of the final round, the *boosted classifier* model combines the weighted votes of each classifier, where the weights are computed based on some functions of the classification accuracies of the results reported by the individual classifier. *Adaptive Boosting* (AdaBoost) is a very popular variant of boosting, and for the classification purpose, we have used this approach. The *boosting* function of the *adabag* library in R is used in the implementation.

***Random Forest***: Random forest is an *ensemble machine learning* approach. The algorithm first builds a large number of decision tree classifiers separately so that the collection of the classifiers is a *forest*. The individual decision tree classifier models are built based on a random selection of attributes at each node. The splitting at each node is done by randomly selecting the feature and the feature value for splitting to introduce as much randomness as possible. In other words, each decision tree depends on the values of a random vector sampled independently, and with the same distribution for all trees in the forest. The objective of introducing so much randomness in building the decision tree models is to avoid overfitting of the models during the training phase. During the classification phase, each tree votes, and the test case is assigned the class that is returned by the majority of the trees. We use the *randomForest* function defined in the *randomForest* library in R in our implementation.

***Artificial Neural Network***: An *artificial neural network* (ANN) is a *connectionist network* that consists of nodes and their interconnecting links where the nodes are arranged in several layers - an input layer, one or more hidden layers, and an output layer. The nodes in the input layer correspond to the predictor variables (i.e., attributes) in the training dataset. The inputs are fed simultaneously into the units making up the input layer. The input values pass through the respective nodes in the input layer and are then weighted using the weights associated with the links connecting the nodes and fed simultaneously to the second layer of nodes, known as the hidden layer nodes. The outputs of the nodes in the first hidden layer are weighted again using the corresponding link weights, and the resultant values are provided as the inputs to a possible second hidden layer and so on. The weighted outputs of the last hidden layer are input to units making up the output layer, which produces the network's prediction for given tuples. We use the *neuralnet* function defined in the *neuralnet* library in R for implementing the ANN classifier. The raw data is normalized using the *min-max normalization* approach. Only the predictors are normalized, the response variable: *open_perc* is kept unchanged. The parameter *hidden* of the function *neuralnet* is varied to change the number of nodes in the hidden layer. The parameter *stepmax* is set to the maximum value of $10^6$ so that the maximum number of iteration capability of the *neuralnet* function can be utilized. For implementing a classification model, the parameter *linear.output* is set to FALSE in the *neuralnet* function.

***Support Vector Machine***: A support vector machine (SVM) model, when applied for classification, can classify both linear and nonlinear data. For linearly inseparable data, it uses a nonlinear mapping to transform the original data into a higher dimension. Within this new higher dimension, it searches for the *linear optimal hyperplane* that separates the two classes. SVM finds this hyperplane using *support vectors* which are the essential and the discriminating training tuples to separate the two classes. We use the *ksvm* function defined in the *kernlab* library in R for carrying implementing the SVM classifier. The function

*ksvm* has an optional parameter called kernel which is set to *vanilladot* in our implementation. The resultant SVM is a linear one as the training data exhibited linear characteristics.

We now briefly discuss the regression models.

**Multivariate Regression**: In this regression approach, we use *open_perc* as the response variable and the remaining ten variables as the predictors to build predictive models for the three cases mentioned earlier in Section 3. In all these cases, we use the programming language R for data management, model construction, testing of models, and visualization of results.

*Case I*: We use 2013 data as the training data set for building the model, and then test the model using the same data set. For both the stocks, we used two approaches of multivariate regression - (i) *backward deletion* and (ii) *forward addition* of variables. Both approaches yielded the same results for the stock price data.

For the year 2013, we apply the *vif* function in the *faraway* library to detect the collinear variables to get rid of the *multicollinearity* problem. The *variance inflation factor* (VIF) values of the variables are found to be as follows: *month* = 1.003, *day_month* = 1.008, *day_week* = 1.002, *time* = 1.095, *high_perc* = 4372.547, *low_perc* = 4369.694, *close_perc* = 165.436, *vol_perc* = 1.072, *nifty_perc* = 1.046, *range_diff* = 156.198. Hence, it is clear that *high_perc*, *low_perc*, *close_perc,* and *range_diff* exhibit multicollinearity. We retain *low_perc* and *range_diff* for the model construction and remove the other two variables since their VIF values are smaller than the other two. Using the *drop1* function in the case of the *backward deletion technique* and the *add1* function in the case of the *forward addition technique*, we identify the variables that are not significant in the model and do not contribute to the information content of the model. For identifying the variables that contribute least to the information contained in the model in each iteration, we use the *Akaike Information Criteria* (AIC) - the variable that has the least AIC value and non-significant *p*-value in each iteration, is removed from the model in case of the backward deletion process. On the other hand, the variable that has the lowest AIC and a significant *p*-value is added to the model in each iteration for the forward addition technique. It is found that *low_perc* and *range_diff* are the two predictors that finally remain in the regression model.

*Case II*: For the year 2014, the VIF values for the predictors are found to be as follows: *month* = 1.007, *day_month* = 1.004, *day_week* = 1.007, *time* = 1.057, *high_perc* = 1161.446, *low_perc* = 1331.035, *close_perc* = 115.161, *vol_perc* = 1.022, *range_diff* = 92.092, *nifty_perc* = 1.073. The variables *high_perc*, *low_perc*, *close_perc*, and *range_diff* exhibit multicollinearity. As in *Case I*, we retain *low_perc* and *range_diff* as their VIF values are smaller compared with the other two. Use of *backward deletion* and *forward addition* methods both yield the same regression models as in *Case I* with *low_perc* and *range_diff* as the predictors and *open_perc* as the response variable.

*Case III*: In this case, the model is identical to that of *Case I*. However, the model is tested on the data of the year 2014. Therefore, the performance results of the model are expected to be different. The performance results and their critical analysis are presented in Section 6.

**Multivariate Adaptive Regression Spline**: *Multivariate Adaptive Regression Spline* (MARS) is a statistical machine learning approach for building robust regression models. MARS works by splitting input variables into multiple *basis functions* and then fitting a linear regression model to those *basis* functions. The basis functions used by MARS are designed in pairs: $f(x) = \{x - t, if\ x > t, 0\ otherwise\}$ and $g(x) = \{t - x, if\ x < t, 0\ otherwise\}$. The main characteristic property of the basis functions is that these functions are *piecewise linear*. The value *t* at which the two functions meet is called a *knot*. The working principles of MARS are very similar to that of CART. Like CART, MARS first builds a complex model involving a large number of basis functions, which are separated from each other by a large number of knots. This phase of the algorithm execution is called the *forward pass* of the model building. In the subsequent phase, known as the *backward pass*, the algorithm prunes back unimportant terms (i.e., basis functions), which could not contribute significantly to the *generalized $R^2$* values of the model. This phase essentially enables MARS to avoid the possibility of overfitting of the model during the training phase. During the execution of the *backward pass*, the algorithm computes the *generalized cross-validation* (GCV) values to determine how well the model fits into the data while avoiding any overfitting. Finally, the algorithm returns the model with the best cost/benefit ratio. To fit a model using MARS in R, we use the function *earth* in the library *earth*.

***Decision Tree***: For building a decision tree regression model, we use the same *tree* function in the *tree* library in R as we did in the decision tree-based classification model. However, in this case, the response variable (i.e., *open_perc*) is a continuous numeric variable. The *predict* function is used to predict the values of the response variable. The functions *cor* and *rmse* defined in the library *Metrics* are used to compute the correlation coefficient and the RMSE value for determining the prediction accuracy of the models.

***Bagging***: We use the *bagging* function defined in the *ipred* library of R for building a bagging regression model. The value of the parameter *nbag* is set to 100 so that one hundred decision trees are ensembled in the regression model. The *predict* function in the *ipred* library is used to predict the response variable values. The *rmse* function in the *Metric* library is used to compute the RMSE of the predicted values. The *cor* function is used to compute the correlation between the actual and the predicted values.

***Boosting***: We use the *blackboost* function defined in the *mboost* library in R for building boosting regression models on the stock price data. As in other regression models, the *predict* and *rmse* functions are used to compute the predicted values and the RMSE values of the regression model.

***Random Forest***: We use the *randomForest* function defined in the *randomForest* library in R for fitting a random forest regression model into the training data. The response variable *open_perc* is a numeric variable, and the same *predict*, *rmse*, and *cor* functions are used as in the case of other regression methods.

***Artificial Neural Network***: As in the case of classification, we use the *neuralnet* function defined in the *neuralnet* library in R for regression on the stock price data. The predictors are normalized using *min-max normalization* before building the model. The *compute* function defined in the *neuralnet* library is used for computing the predicted values, while the parameter *hidden* is used to change the number of nodes in the hidden layer. The value of the parameter *stepmax* is set to $10^6$ to exploit the maximum number of iterations executed by the *neuralnet* function. The parameter *linear.output* is by default set to TRUE, and hence it is not altered. For the Godrej Consumer Products training dataset, only one node in the hidden layer for all the three cases was sufficient for building ANN regression models.

***Support Vector Machine***: For building the SVM regression model, we use the *svm* function defined in the *e1071* library in R. The *predict* function is used for predicting the response variable values using the regression model, and the *rmse* function is used to compute the RMSE values for the predicted quantities.

## 5. Deep Learning Models

In this section, we discuss two deep learning-based regression methods: (i) the *long- and short-term memory* (LSTM) network, and (ii) the *convolutional neural networks* (CNNs).

**Long- and Short-Term Memory Network**: LSTM is a variant of *recurrent neural networks* (RNNs) - neural networks with feedback loops (Geron, 2019). In such networks, output at the current time slot depends on the current inputs as well as the previous state of the network. However, RNNs suffer from the problem that these networks cannot capture long-term dependencies due to *vanishing or exploding gradients* during backpropagation in learning the weights of the links (Geron, 2019). LSTM networks overcome such problems, and hence such networks are quite effective in forecasting in multivariate time series. LSTM networks consist of memory cells that can maintain their states over time using memory and gating units that regulate the information flow into and out of the memory. There are different variants of gates used. The forget gates control what information to throw away from memory. The input gates are meant for controlling the new information that is added to the cell state from the current input. The cell state vector aggregates the two components - the old memory from the forget gate, and the new memory from the input gate. In the end, the output gates conditionally decide what to output from the memory cells. The architecture of an LSTM network along with the *backpropagation through time* (BPTT) algorithm for learning provides such networks a very powerful ability to learn and forecast in a multivariate time series framework. We use Python programming language and the Tensorflow and Keras deep learning frameworks for implementing LSTM networks. While building the LSTM models, we use the *open* price of the stock as the response variable, and the variables high, low, close, volume and NIFTY, are used as the predictors. Unlike the machine learning techniques, for the LSTM models, we don't compute the differences between successive slots. Rather, we forecast the *open* value of the next slot based on the values of the response and the predictor variables in the previous slots.

We use the *mean absolute error* (MAE) as the *loss function* and the *adaptive moment estimation* (ADAM) as the *optimizer* for evaluating the model performance in all three cases. ADAM computes adaptive learning rates for each parameter in the gradient descent algorithm. In addition to storing an exponentially decaying average of the past squared gradients, ADAM also keeps track of the exponentially decaying average of the past gradients, which serves as the momentum in the learning process. Instead of behaving like a ball running down a steep slope like momentum, ADAM manifests itself like a heavy ball with a rough outer surface. This high level of friction results in ADAM's preference for a flat minimum in the error surface. Due to its ability to integrate adaptive learning with a momentum, ADAM is found to perform very efficiently in optimizing the performance of large-scale networks. This was the reason for our choice of ADAM as the optimizer in our LSTM modeling.

We train the LSTM networks using different epoch values and batch sizes for the three different cases. The *sequential constructor* in the *Tensorflow* framework is used in building the LSTM model. The performance results of the LSTM models are presented in Section 6.

**Convolutional Neural Networks**: CNNs emerged from the study of the brain's visual cortex, and they have been used in image recognition since the 1980s. In the last few years, thanks to the increase in computational power, the amount of available training data, and the tricks for training deep neural networks. CNNs have managed to achieve superhuman performance on some complex visual tasks. They power image search services, self-driving cars, automatic video classification systems, and more. Moreover, CNNs are not restricted to visual perception: they are also successful at many other tasks, such as voice recognition, natural language processing, and complex time series analysis of financial data (Binkowski et al., 2017; Lahmiri, 2014).

In the present work, we exploit the power of CNN in forecasting the univariate and multivariate time series data of Godrej Consumer Products stock. CNNs have two major types of processing layers – *convolutional layers* and *pooling or subsampling layers*. The convolutional layers read an input such as a 2-dimensional image or a one-dimensional signal using a kernel (also referred to as the filter) by reading the data in small segments at a time, and scan across the input data field. Each read result is an interpretation of the input that is projected onto a filter map and represents an interpretation of the input. The pooling or the subsampling layers take the feature map projections and distill them to the most essential elements, such as using a signal averaging (average pool) or signal maximizing process (max pool). The convolution and pooling layers are repeated at depth, providing multiple layers of abstraction of the input signals. The output of the final pooling layer is fed into one or more fully-connected layers that interpret what has been read and maps this internal representation to a class value.

We use the power of CNN in multi-step time series forecasting in the following way. The convolutional layers are used to read sequences of the input data, and automatically extract features. The pooling layers are used for distilling the extracted features, and in focusing attention on the most salient elements. The fully connected layers are deployed to interpret the internal representation and output a vector representing multiple time steps. The benefits that CNN provides in our time series forecasting job are the automatic feature learning, and the ability of the model to output a multi-step vector directly.

We build three different types of CNN models for multi-step time series forecasting of stock prices. They are: (i) Multi-step time series forecasting with univariate input data, (ii) Multi-step time series forecasting with multivariate input data via channels – in this case, each input sequence is read as a separate channel, like different channels of an image (e.g., red, green, and blue), (iii) multi-step time series forecasting with multivariate input data via sub-models – in this case, each input sequence is read by a different CNN sub-model and the internal representations are combined before being interpreted and used to make a prediction.

In the first case, we design a CNN for multi-step time series forecasting using only the univariate sequence of the *open* values. In other words, given some number of prior days of *open* values, the model predicts the next standard week of stock market operation. A standard week consists of five days – Monday to Friday. The number of prior days used as the input defines the one-dimensional (1D) data of *open* values that CNN will read and learn for extracting features.

The multi-step time series forecasting approach is essentially an autoregression process. Whether univariate or multivariate, the prior time series data is used for forecasting the values for the next week.

## 6. Performance Results and Analysis

In this Section, we provide a detailed discussion on the implementation, training, and testing of the predictive models and their performance results. We first discuss the classification techniques and then the regression techniques. The metrics that have been used in evaluating the classification models are as follows:

***Sensitivity*:** It is the ratio of the number of true positives to the total number of positives in the test dataset, expressed as a percentage. Here, positive refers to the cases that belong to the target group (i.e., the class

"1"). The term *true positive* refers to the number of positive cases that the model correctly identified. The word *sensitivity* is also sometimes referred to as *recall*.

**Specificity**: It is the ratio of the number of *true negatives* to the total number of negatives in the test dataset, expressed as a percentage. Here, *negative* refers to the cases that belong to the non-target group (i.e., the class "0"). The term *true negative* refers to the number of negative cases that the model correctly classified.

**Positive Predictive Value**: *Positive predictive value* (PPV), also sometimes referred to as *precision*, refers to the accuracy of the model in classifying the target group cases among the total number of target group cases identified by it. It is computed as the ratio of the number of correctly classified target group cases to the total number of target group cases as identified by the model. Since the total number of target group cases identified by the model is the sum of the number of *true positive* cases and the number of *false-positive* cases, PPV is the ratio of the total number of *true positive* cases to the sum of the number of *true positive* cases and the number of *false-positive* cases, expressed as a percentage. The complement of PPV is also called the *false discovery rate* (FDR).

**Negative Predictive Value**: *Negative predictive value* (NPV) refers to the accuracy of the model in classifying the non-target group cases among the total number of non-target elements identified by it. NPV is computed as the ratio of the number of correctly identified non-target group cases to the total number of non-target group cases as identified by the model. Since the total number of non-target group cases classified by the model is the sum of the number of *true negative* cases and the number of *false-negative* cases, NPV is the ratio of the total number of *true negative* cases to the sum of the number of *true negative* cases and the number of *false-negative* cases, expressed as a percentage. The complement of NPV is also called the *false-omission rate* (FOR).

**Classification Accuracy (CA):** It is the ratio of the total number of cases that are correctly classified to the total number of cases in the dataset, expressed as a percentage.

*F1 Score*: If the test data set is highly *imbalanced*, with the cases belonging to the non-target group far outnumbering the number of target group cases, sensitivity is usually found to be very poor even when the *classification accuracy* may be high. Hence, classification accuracy is not considered a very robust and reliable metric. *F1 score*, which is computed as the *harmonic mean* of the *sensitivity* and PPV, is found to be a very robust metric, however.

*Classification Methods*:

**Logistic Regression**: We use *glm* function in the R programming language to build logistic regression-based classification models. The response variable (i.e., *open_perc*) is converted into a categorical type by using the function *as.factor* before building the models. The parameter *family* is set to *binomial* to build a *binary* logistic regression model. The *predict* function is used to predict the class of the test data records. We also build the *lift curve* and the *receiver operating characteristic* (ROC) curve of the model for each case. The output of the *performance* function defined in the ROCR library is plotted to depict the ROC curve of the model. The *area under the curve* (AUC) for each ROC curve is computed using the *auc* function defined in the *pROC* library in the R programming language.

Table 1: Logistic regression classification results

|  | Case I | Case II | Case III |
| --- | --- | --- | --- |

| Metrics | Training Accuracy 2013 | Training Accuracy 2014 | Test Accuracy 2014 |
|---|---|---|---|
| Sensitivity | 94.79 | 94.83 | 92.10 |
| Specificity | 97.61 | 95.96 | 89.39 |
| PPV | 96.87 | 95.12 | 87.83 |
| NPV | 96.01 | 95.72 | 93.16 |
| CA | 96.38 | 95.45 | 90.62 |
| F1 Score | 95.82 | 94.97 | 89.91 |

Table 1 presents the performance results of the logistic regression classification method. For *Case I*, out of 419 actual "0" cases, only 10 cases are misclassified as "1", while among 326 actual "1" cases, 17 cases are found to be wrongly classified as "0". The value of AUC for the ROC curve for *Case I* is 0.9934. For *Case II*, 16 cases out of total 396 actual "0" cases are misclassified as "1", and out of 329 cases which are actually "1", are wrongly classified as "0". The AUC value for this case is found to be 0.9891. *Case III* yields 42 cases which are actually "0" but misclassified as "1" out of a total of 396 cases, while among 329 cases which are actually "1", 26 cases are misclassified as "0". The AUC value for *Case III* is found to be 0.9587.

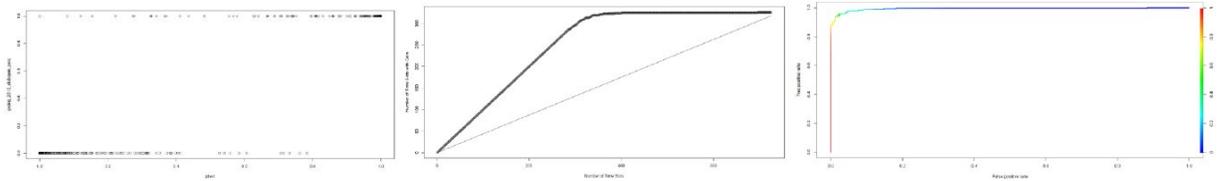

Figure 1: Logistic regression for *Case I* – (a) *actual* vs *predicted* probabilities of *open_perc*, (b) the *lift curve*, and (c) the *ROC curve*

Fig. 1 presents the classification performance, the *lift curve,* and the ROC curve of the logistic regression-based classification model for *Case I*. In Fig. 1(a), the *y*-axis represents the *actual* classes of the records (either "0" or "1") and the *x*-axis denotes the probability that a case will belong to the class "1". The *threshold value* along the *x*-axis is by convention taken to be 0.5. Hence, all the cases which are found to be lying on the level "0" along the *y*-axis and situated to the right of the threshold value of 0.5 along the *x*-axis are misclassified. Similarly, all the points which are on the level "1" along the *y*-axis, and are situated to the left of the threshold value of 0.5 along the *x*-axis are also misclassified. It is evident from Fig. 1(a) that the number of misclassified cases in *Case I* is very low. Fig. 1(b) shows that the *lift curve* is pulled up from the baseline indicating that the model is very effective in discriminating between the two classes. Fig. 1(c) depicts the ROC curve for the logistic regression model for *Case I*. The steepness of the curve makes it evident that the model can very effectively optimize the values of the *true positive rate* (TPR) and the *false positive rate* (FPR). In Fig. 1(c), the line segment with *red* color presents the class "1" cases that are correctly classified, while the *blue* line segment denotes the correctly classified cases which belong to the class "0". The portion of the ROC curve that is colored with *yellow* represents those cases that belong to the class "0", but the model wrongly classified them to the class "1". The *green*-colored portion of the ROC curve depicts those cases which are misclassified into the class "0", while they belong to the class "1".

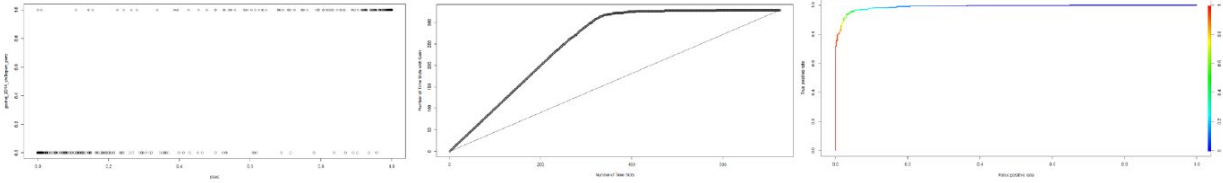

Figure 2: Logistic regression for *Case II* – (a) *actual* vs *predicted* probabilities of *open_perc*, (b) the *lift curve*, and (c) the *ROC curve*

Fig. 2 depicts respectively the *classification performance*, the *lift curve*, and the ROC curve of the logistic regression model for *Case II*. The performance of the model, in this case, is similar to that in *Case I*. However, the AUC value yielded by the model under this case is just marginally smaller than the corresponding value in *Case I*.

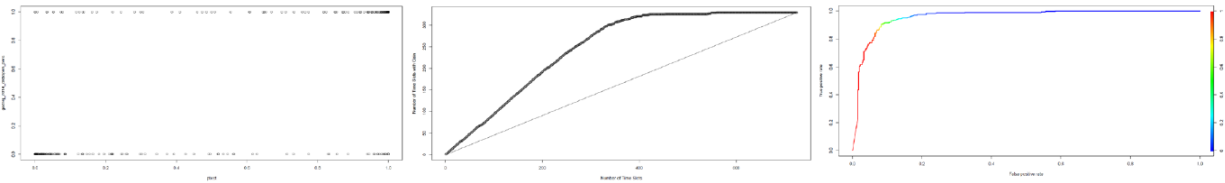

Figure 3: Logistic regression for *Case III* – (a) *actual* vs *predicted* probabilities of *open_perc*, (b) the *lift curve*, and (c) the *ROC curve*

Fig. 3 shows the classification accuracy, the lift curve, and the ROC curve for the logistic regression model in *Case III*. It is evident from Fig. 3(c) that the classification model in *Case III* has committed more errors as compared to *Case I* and *Case II*. This case also yields a lower AUC value of 0.9587.

*KNN Classification*: Table 2 presents the performance results of the KNN classification method. For *Case I*, with the values of $k = 1, 3, 5, 7$, and 9, the classification accuracy values are found to be 100, 93.42, 91.68, 92.35, and 92.08 respectively. We choose $k = 3$ to avoid the overfitted model with $k = 1$. In this case, 419 cases are 0s and 326 cases are 1s. 15 cases of actual 0s are misclassified as 1s, and 34 cases of actual 1s are misclassified as 0s. In *Case II*, for $k = 1, 3, 5, 7$, and 9, the classification accuracy values are 100, 90.21, 85.10, 83.22, and 84.16 respectively. Again $k = 3$ is chosen to avoid model overfitting. 28 cases of actual 0 are misclassified as 1, while 43 cases of actual 1 are misclassified as 0. For *Case III*, the classification accuracy values are found to be 65.24, 65.10, 67.17, 68.69, and 67.44 for $k = 1, 3, 5, 7$, and 9 respectively. We choose $k = 3$, for which 202 cases which are 0s, are misclassified as 1s, while 51 cases of actual 1s were misclassified as 0s.

Table 2: KNN classification results

| Metrics | Case I<br>Training Accuracy 2013 | Case II<br>Training Accuracy 2014 | Case III<br>Test Accuracy 2014 |
|---|---|---|---|
| Sensitivity | 89.57 | 86.93 | 84.50 |
| Specificity | 96.42 | 92.93 | 48.99 |
| PPV | 95.11 | 91.08 | 57.92 |
| NPV | 92.24 | 89.54 | 79.18 |
| CA | 93.42 | 90.21 | 65.10 |
| F1 Score | 92.26 | 88.96 | 68.73 |

***Decision Tree Classification*:** We use the *tree* function defined in the *tree* library in the R programming language for building the decision tree-based classification models. The response variable *open_perc* is converted into a categorical type using the *as.factor* function for classification. The *predict* function in the *tree* library is used for predicting the classes of the response variable for the records in the test dataset. For *Case I* and *Case III* the models are identical as they are trained on the same training dataset of the year 2013. However, while the model in *Case I* is tested on the 2103 data, the 2014 data is used for testing the model in *Case II*. For all three cases, we find that *high_perc*, *low_perc*, and *close_perc* are the three predictor variables that are used in constructing the models. However, in *Case I*, the predictor which is used for splitting at the root node is *close_prec*, indicating that *close_perc* is the most important predictor for classification in the 2013 dataset. However, for the 2014 dataset, *high_perc* is found to be the most discriminating one as the same is used by the model for splitting at the root node. In *Case I*, the decision tree classifier misclassifies 8 cases out of a total of 419 cases which belong to the class "0", while 16 cases are misclassified out of a total of 326 cases which are the records of the class "1".

Table 3: Decision Tree classification results

| Metrics | Case I<br>Training Accuracy 2013 | Case II<br>Training Accuracy 2014 | Case III<br>Test Accuracy 2014 |
|---|---|---|---|
| Sensitivity | 95.09 | 92.40 | 89.97 |
| Specificity | 98.09 | 95.71 | 92.42 |
| PPV | 97.48 | 94.70 | 90.80 |
| NPV | 96.25 | 93.81 | 91.73 |
| CA | 96.78 | 94.21 | 91.31 |
| F1 Score | 96.27 | 93.54 | 90.38 |

In *Case II*, the model fails to classify correctly 17 cases out of a total of 396 cases which are actually "0" class members, while 25 cases are misclassified out of a total of 329 cases that belong to the class "1". *Case III*, faces a more challenging task. We find that the model is unable to correctly classify 30 cases out of a total of 396 cases that belong to the class "0", while 33 cases are misclassified out of a total of 329 cases that belong to the actual class of "1". Table 3 presents the performance results of the decision tree classification models under three different cases. Fig. 4 depicts the decision tree classifiers for *Case I*, *Case II*, and *Case III* respectively.

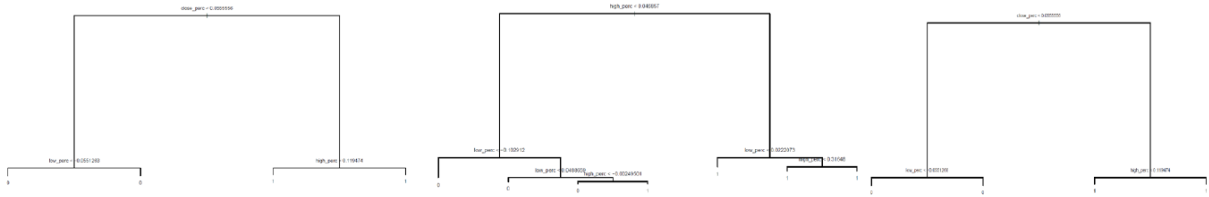

Figure 4: Decision tree for classification – (a) tree for *Case I*, (b) the tree for *Case II*, and (c) the tree for *Case III*

Table 4: Bagging classification results

| Metrics | Case I<br>Training Accuracy 2013 | Case II<br>Training Accuracy 2014 | Case III<br>Test Accuracy 2014 |
|---|---|---|---|
| Sensitivity | 95.09 | 95.44 | 89.97 |
| Specificity | 98.09 | 96.46 | 92.42 |
| PPV | 97.48 | 95.73 | 90.78 |
| NPV | 96.25 | 96.22 | 91.73 |
| CA | 96.78 | 96.00 | 91.31 |
| F1 Score | 96.07 | 95.58 | 90.37 |

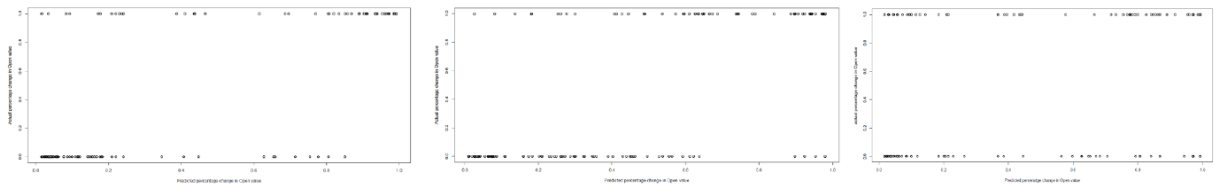

Figure 5: Bagging classification: *actual* vs *predicted open_perc* – (a) for *Case I*, (b) for *Case II*, and (c) for *Case III*

***Bagging Classification***: We used the *bagging* function defined in the *ipred* library in the R programming language for building the bagging classification models. We set the value of the parameter *nbag* to 25 so that 25 decision trees were created randomly, and a simple majority voting mechanism is applied in constructing the classifier. In *Case I*, we find that the model fails to correctly classify 8 cases out of a total of 419 cases that belong to the class "0", while 16 cases out of a total of 326 cases that belong to the class "1" are also misclassified. In *Case II*, the model is unable to correctly classify 14 cases out of a total of 396 cases that are of actual class "0", while 15 cases out of a total of 329 cases are misclassified which belong to the class "1". In *Case III*, 30 cases out of 396 actual "0" cases are misclassified by the model, while 33 cases out of a total of 329 cases of the class "1" are also misclassified. The performance results of the bagging classification model for all three cases are presented in Table 4. Fig. 5 depicts the classification accuracy of the model for *Case I*, *Case II*, and *Case III* respectively. In all these three figures, the *y*-axis represents the *actual* class labels, while the values along the *x*-axis show the probabilities of the predicted class for the records. The cases which are on the label "0" on the *y*-axis and have their probability values greater than 0.5 along the *x*-axis are the misclassified cases. In a similar line, those cases which are lying on the label "1" along the *y*-axis and have their probability values less than 0.5 along the *x*-axis, are also misclassified.

Table 5: Boosting classification results

| Metrics | Case I | Case II | Case III |
|---|---|---|---|
| | Training Accuracy 2013 | Training Accuracy 2014 | Test Accuracy 2014 |
| Sensitivity | 100 | 100 | 92.10 |
| Specificity | 100 | 100 | 93.43 |
| PPV | 100 | 100 | 92.10 |
| NPV | 100 | 100 | 93.43 |
| CA | 100 | 100 | 92.83 |
| F1 Score | 100 | 100 | 92.10 |

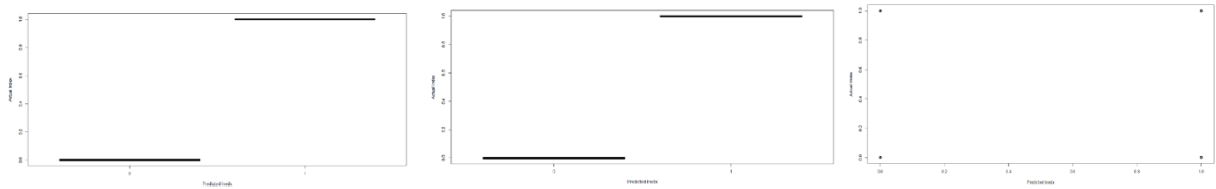

Figure 6: Boosting classification: *actual* vs predicted open_perc – (a) for *Case I*, (b) for *Case II*, and (c) for *Case III*

**Boosting Classification**: We use the *boosting* function defined in the *adabag* library in the R programming language for building the boosting models for classification. The response variable *open_perc* is transformed into the categorical type using *as.factor* function to satisfy the requirement of a classification model. The *predict* function is used for predicting the class of the response variable in the test data records. For both *Case I* and *Case II*, the boosting classification models are found to be yielding 100% accuracy in all the metrics of classification. This may be observed in Table 5. This is not surprising as in both the cases the models were built and tested using the same dataset, and thus the learning of the models had been very accurate using the ensemble of the *weighted majority voting* on a large number of random decision tree classifiers. However, the model encounters a more difficult task in *Case III*, in which the ensemble model is built on the 2013 data, and then tested on the 2014 data. In *Case III*, we find that the model misclassifies 26 cases out of a total of 396 cases that belong to the class "0", while among 329 cases that are actually of the class "1", 26 cases are incorrectly classified. Table 5 presents the performance results of the boosting classification models. Fig. 6 depicts the performance of the boosting classifier for *Case I*, *Case II*, and *Case III* respectively. In these three figures, along the *y*-axis the *actual* classes are plotted – there are two actual class levels "0" and "1". The *x*-axis presents the predicted probability that a case will belong to the class "1". Hence, the data points which are situated to the left side of the threshold value of 0.5 along the *x*-axis, and lying on the level "1" along the *y*-axis are the misclassified cases. Similarly, the points that are on the right side of the threshold value of 0.5, and lying on the level "0" along the *y*-axis are also the misclassified cases. It is evident from Fig. 6 that boosting classifiers have performed very well in all three cases.

Table 6: Random Forest classification results

| Metrics | Case I<br>Training Accuracy 2013 | Case II<br>Training Accuracy 2014 | Case III<br>Test Accuracy 2014 |
|---|---|---|---|
| Sensitivity | 94.48 | 93.01 | 91.19 |
| Specificity | 97.61 | 94.19 | 92.93 |
| PPV | 96.86 | 93.01 | 91.46 |
| NPV | 98.08 | 94.19 | 92.70 |
| CA | 96.24 | 93.66 | 92.14 |
| F1 Score | 95.66 | 93.01 | 91.32 |

***Random Forest Classification***: We use the *randomForest* function defined in the *randomForest* library in R programming language, for building random forest-based classification models. In all three cases, the random forest algorithm creates 500 decision trees using three predictors at each node in the decision trees for carrying out the splitting task. In *Case I*, the model misclassifies 10 cases as the class "1" cases out of a total of 419 cases that belong to the class "0". On the other hand, 18 cases are misclassified into the class "1" out of a total of 326 cases which are actually of the class "0". The *out-of-bag* (OOB) estimate of the error rate of the model, in this case, is 3.76%. In *Case II*, the model fails to classify correctly 23 cases out of a total of 396 cases that belong to the actual class of "0". On the other hand, 23 cases out of a total of 329 cases that belong to the class of "0" are also misclassified. The OOB estimate of the error rate of the classification model, in this case, is 6.34%. In Case III, the random forest classification model was identical to that in Case I. However, the model was tested on 2014 data unlike the model in Case I that was tested on 2013 data. We find that in *Case III*, the model misclassifies 28 cases out of a total number of 396 cases that belong to the class "0". On the other hand, 29 cases are wrongly classified out of a total of 329 actual "1" cases. The performance results of the random forest classification models are presented in Table 6.

***ANN Classification***: We use the *neuralnet* function defined in the *neuralnet* library in the R programming language to build ANN classification models. The parameter *linear.output* is set to *false*, and the response variable *open_perc* is converted into a categorical variable type by using the function *as.factor* before the classification models are built. We find that only one node at the hidden layer is sufficient to model the data. Hence, we pass the value of the parameter *hidden* as 1 in the *neuralnet* function. To avoid any possible scenario in which the *backpropagation* algorithm fails to converge, we set the parameter *stepmax* to its maximum possible value of $10^6$. In *Case I*, the ANN classification model misclassifies 10 cases out of a total of 419 cases as "1", while they belong to the class "0". On the other hand, 15 cases which are actually "1", are misclassified as "0" out of a total of 326 cases. The ANN model for classification for *Case I* and its performance in the classification task is presented in Fig. 7. In Fig. 7(b), the actual class labels are plotted along the *y*-axis, and along the *x*-axis, the predicted probabilities are plotted. The points lying on the actual class label "0" along the *y*-axis while having their predicted class probabilities greater than 0.5 (i.e., those points on the "0" label lying on the right-hand side of the threshold value of 0.5 along the x-axis) represent the misclassified cases. In a similar line, the points which are on the label "1" along the *y*-axis while having their probabilities smaller than 0.5 (i.e., those points on the "1" label lying on the left-hand side of the threshold value of 0.5 along the x-axis) are also misclassified points. In *Case II*, the ANN classification model misclassifies 17 cases as class "0" out of 396 cases that belong to the class "1". On the other hand, 21 cases are misclassified as class "1" out of 329 cases which are class "0" cases.

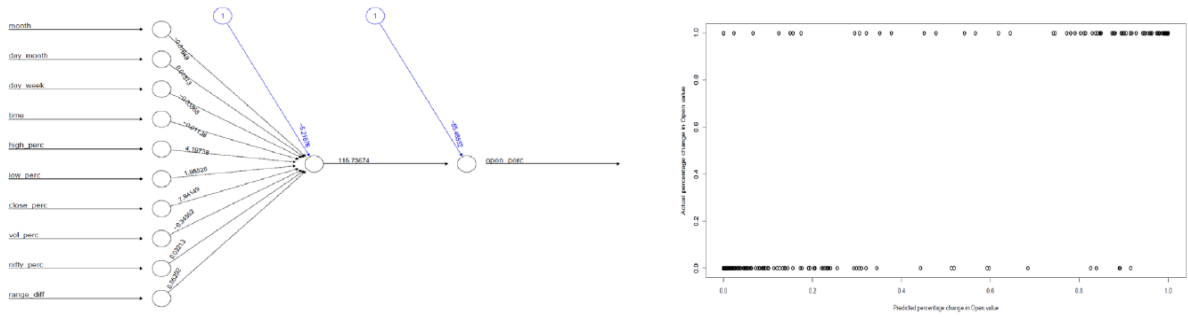

Figure 7: ANN classification – (a) the model for *Case I*, (b) *actual* vs *predicted open_perc* for *Case I*

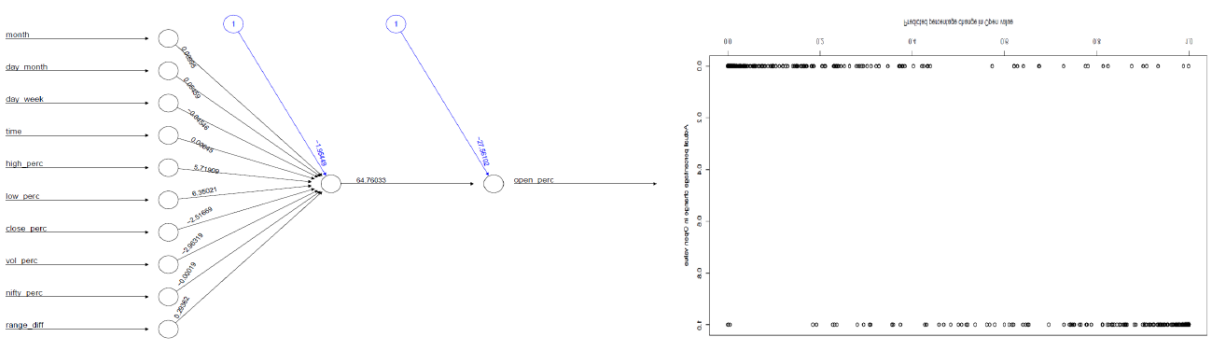

Figure 8: ANN classification – (a) the model for *Case II*, (b) *actual* vs *predicted open_perc* for *Case II*

Fig. 8 presents the ANN classification model in *Case II*, and its performance in the classification task, respectively. In *Case III*, the model is built using 2013 data, hence it is identical to the model that is used in *Case I*. However, since the model is tested on 2014 data, unlike in *Case I* in which the model is tested on 2013 data, the performance results of the model in *Case III* are quite different. The model in *Case III* encounters a more difficult challenge as there are differences in the characteristics of the data in 2013 and 2014. We find that in *Case III*, the model misclassifies 259 cases as class "1" out of 396 cases that belong to the class "0". On the other hand, only 1 case out of 329 cases which are actually of the class "1" are misclassified as the class "0". It is evident, that model performs poorly in classifying the class "0" cases which results in a very low value of its specificity. The specificity in *Case III* is found to be only 34.60%, while for *Case I* and *Case II*, the specificity values are 97.61% and 95.71% respectively. This indicates that the ANN classification model has a poor generalization in its learning during the training phase using the 2013 data, and that possibly led to a model overfitting. The overfitted model fails to correctly classify the majority of the "0" cases in the test data of 2014. This results in a very low specificity value. Fig. 9 presents the ANN classification model, and its classification performance respectively. Table 7 presents the performance of the ANN classification models.

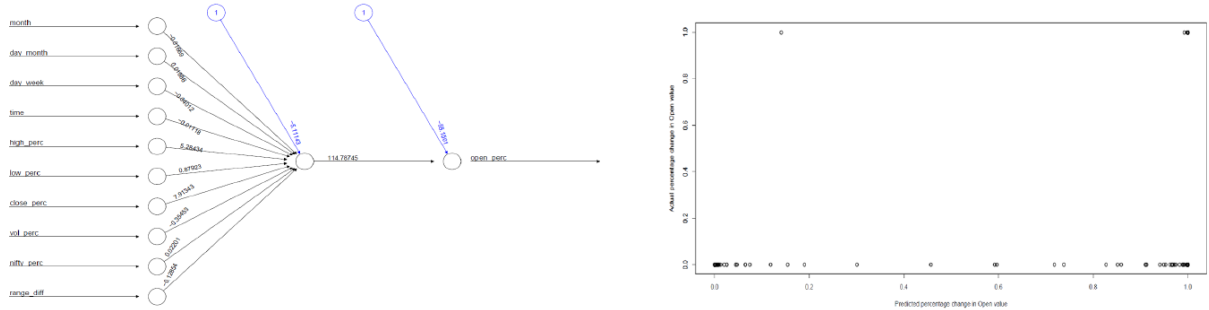

Figure 9: ANN classification – (a) the model for *Case III*, (b) *actual* vs *predicted open_perc* for *Case III*

Table 7: ANN classification results

| Metrics | Case I | Case II | Case III |
|---|---|---|---|
|  | **Training Accuracy 2013** | **Training Accuracy 2014** | **Test Accuracy 2014** |
| Sensitivity | 95.40 | 93.62 | 99.70 |
| Specificity | 97.61 | 95.71 | 34.60 |
| PPV | 96.88 | 94.77 | 55.88 |
| NPV | 96.46 | 94.75 | 99.28 |
| CA | 96.64 | 94.76 | 64.14 |
| F1 Score | 96.13 | 94.19 | 71.62 |

***SVM Classification*:** We use the *ksvm* function defined in the *kernlab* library in the R programming language for building the SVM-based classification models. The function *ksvm* is used with the parameter *kernel* set to *vaniladot*. It implies that a *linear kernel* is used for building the SVM classification models. For *Case I*, the model finds 120 *support vectors*. We observe that out of a total number of 430 cases which are actually "0" class records, 19 cases are misclassified as "1". On the other hand, 8 cases are misclassified as "0", out of a total of 315 cases which are actually "1". The *training error* in *Case I* is found to be 3.62%. For *Case II*, the model detects 156 support vectors for classifying 725 records. Among 406 cases that belong to the class "0", 27 cases are misclassified as "1". On the other hand, 17 cases are misclassified as "0" out of a total of 319 cases which are actually "1". The *training error* for *Case II* is found to be 6.07%. The SVM classification model identifies 116 support vector points in *Case III*. The model misclassified 41 cases as "1" out of a total of 418 cases which are actually "0". On the other hand, out of a total of 307 cases which are actually "1", 19 cases are misclassified as "0". Table 8 presents the results of the SVM classification models.

Table 8: SVM classification results

| Metrics | Case I | Case II | Case III |
|---|---|---|---|
| | TrainingAccuracy 2013 | Training Accuracy 2014 | Test Accuracy 2014 |
| Sensitivity | 94.46 | 94.67 | 93.81 |
| Specificity | 95.58 | 93.35 | 90.19 |
| PPV | 94.17 | 91.79 | 87.54 |
| NPV | 98.09 | 95.71 | 95.20 |
| CA | 96.38 | 93.93 | 91.72 |
| F1 Score | 94.31 | 93.21 | 90.57 |

*Regression Methods*:

*Multivariate Regression*: We already mentioned in Section 4 4 that the predictors that are finally included in the *multivariate regression* models in all three cases are: *low_perc* and *range_diff*. For *Case I*, the regression model yields a value of 0.9919 for the *adjusted $R^2$* value, and the *F* statistic value of $4.58*10^4$ with an associated *p*-value of $2.2*10^{-16}$. This indicates that the regression model is successfully able to establish a linear relationship between the response variable, *open_perc*, and the predictor variables, *low_perc*, and *range_diff*. The RMSE value yielded by the regression model for this case is found to be 0.0853, and the mean of the absolute values of the actual *open_perc* is 0.6402. The ratio of the RMSE to the mean of the absolute values of the actual *open_perc* is found to be 13.317. 14 cases out of a total of 745 cases exhibit sign mismatch between the *predicted* and the *actual* values of *open_perc*. The correlation test produces the value of the correlation coefficient value as 0.99 with a *p*-value of the *t*-statistic as $2.2*10^{-16}$. This indicates that there is a strong *linear* relationship between the predicted and the actual values of *open_perc*. The *Breusch-Pagan* test for homoscedasticity on the residuals yields a test statistic value of 10.239 with a *p*-value of 0.005978. Hence, it is evident that the residuals are not *homoscedastic*. However, the *Durbin-Watson* test of autocorrelation produces a test statistic value of 3.023 with an associated *p*-value of 1. Hence, the null hypothesis that assumes the presence of no autocorrelation among the residuals has the fullest support. Hence, we conclude that the *residuals do not exhibit any significant autocorrelation*. For *Case II*, the regression model yields an *adjusted $R^2$* value of 0.9827 with the value of the *F*-statistics as $2.052*10^4$. The *p*-value of the *F* statistics is found to be smaller than $2.2*10^{-16}$ indicating a very *highly significant F* statistic, and a very good model fit. The RMSE value for *Case II* is found to be 0.1749, with the mean of the absolute values of the *actual open_perc* as 0.9286. The ratio of the RMSE to the mean of the absolute values of the *actual open_perc* is found to be 18.84. 39 cases out of a total of 725 cases have a *sign mismatch* between the predicted and the actual *open_perc* values. The correlation test for this case yields a value of correlation coefficient as 0.99 with a value of the *t*-statistic as 202.74. The *p*-value of the *t*-statistic is $2.2*10^{-16}$, which indicates a very strong linear relationship between the *predicted* and the *actual open_perc* values. The *Breusch-Pagan* test yields a test statistic value of 3.1877 with an associated *p*-value of 0.203. Hence, it is evident that the residuals do not exhibit the presence of any significant heteroscedasticity. The *Durbin-Watson* test of autocorrelation produces a test statistic value of 2.9005. The *p*-value of the *Durbin-Watson* test is found to be 1. This indicates that the null hypothesis of *no significant autocorrelation among the residues* has received full support. Hence, we conclude that the residuals in the regression model for *Case II* do not exhibit any significant autocorrelation. The model built under *Case III* is the same as the one built under *Case I*. However, its performance results are quite different as it is tested on 2014 data, unlike the model in *Case I* which is tested on 2013 data. The RMSE of the model for *Case III* is found to be 0.1753, with the mean of the absolute values of the *actual open_perc* as 0.9286. Thus, the ratio of the RMSE to the mean of the absolute values of the *actual open_perc* values is found to be 18.88. We find that 39 cases out of a total of 725 cases exhibit a *sign mismatch* between the *predicted* and the *actual* values of *open_perc*. The correlation test yields a correlation coefficient of 0.99, with the value of the *t*-statistics as 202.53, and the associated *p*-value of $2.2*10^{-16}$. This indicates that the *predicted* and the *actual* values of *open_perc* exhibit a strong linear relationship between them. The *Breusch-Pagan* test yields a test statistic value of 3.1877 with an associated *p*-value of 0.2031. Hence, it is clear that the residuals are not *heteroscedastic*. The test statistic value yielded by the *Durbin-Watson* test is found to be 2.9005 with an associated *p*-value of 1. Thus, it is evident that the null hypothesis of no autocorrelation among the residuals has gained the fullest support. We conclude that the residuals do not exhibit any significant autocorrelation.

Table 9: Multivariate Regression results

| Metrics | Case I<br>Training 2013 | Case II<br>Training 2014 | Case III<br>Test 2014 |
| --- | --- | --- | --- |
| Correlation Coefficient | 0.99 | 0.99 | 0.99 |
| RMSE/Mean of Absolute Values of Actuals | 13.32 | 18.84 | 18.88 |
| Percentage of Mismatched Cases | 18.67 | 5.38 | 5.24 |

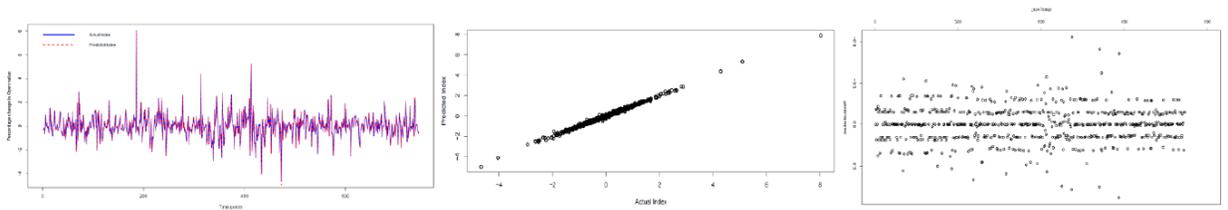

Figure 10: Multivariate regression – (a) *actual* and *predicted* values of *open_perc* for *Case I*, (b) relationship between *actual* and *predicted open_perc* for *Case I*, (c) *residuals* for *Case I*

Table 9 presents the performance results of the multivariate regression models. Fig. 10 presents the performance results of the model under *Case I*. It is evident from Fig. 10(a) that the *predicted* values very closely follow the pattern of the variation of the *actual open_perc* values. Fig. 10(b) exhibits a very strong linear relationship between the *predicted* and the *actual* values of *open_perc*. Fig. 10(c) shows that the residuals of the model are scattered and random, and they do not exhibit any significant autocorrelation.

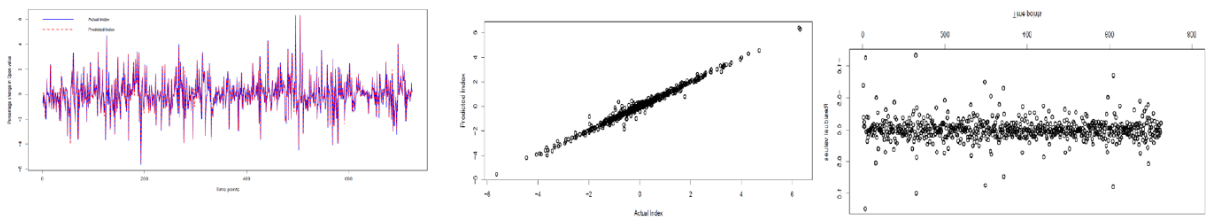

Figure 11: Multivariate regression – (a) *actual* and *predicted* values of *open_perc* for *Case II*, (b) relationship between *actual* and *predicted open_perc* for *Case II*, (c) *residuals* for *Case II*

The performance results of *Case II* are presented in Fig. 11. The *predicted* and the *actual* values of the *open_perc* exhibit almost identical movement patterns in *Case II* as in *Case I*. The residuals do not show any significant autocorrelations. The performance results of the model for *Case III* are presented in Fig. 12. Fig. 12(a) shows how closely the predicted values of the *open_perc* follow the patterns exhibited its *actual* values in *Case III*, while Fig 12(b) exhibits a strong linear relationship between them. Fig. 12(c) shows that the residuals in *Case III* are random, and do not exhibit any significant autocorrelations.

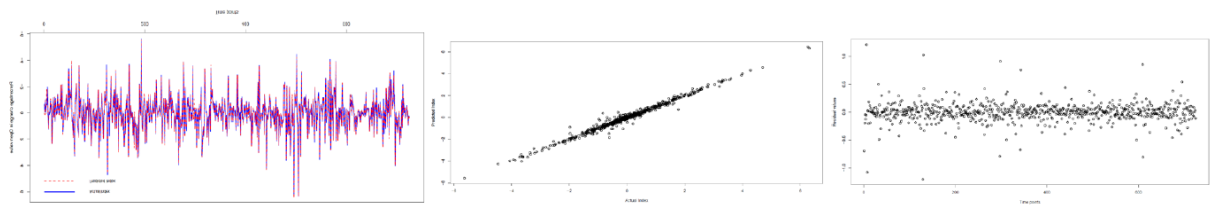

Figure 12: Multivariate regression – (a) *actual* and *predicted* values of *open_perc* for *Case III*, (b) the relationship between *actual* and *predicted open_perc* for *Case III*, (c) *residuals* for *Case III*

Table 10: MARS regression results

|  | Case I | Case II | Case III |
|---|---|---|---|
| **Metrics** | Training 2013 | Training 2014 | Test 2014 |
| Correlation Coefficient | 0.99 | 0.99 | 0.99 |
| RMSE/Mean of Absolute Values of Actuals | 12.41 | 17.09 | 20.40 |
| Percentage of Mismatched Cases | 1.21 | 4.28 | 6.34 |

*Multivariate Adaptive Regression Spline* (*MARS*)**:** We use the *earth* function defined in the *earth* library in the R programming language for building the MARS regression models. For *Case I*, in the *forward pass* of the execution of the algorithm, seven terms are used in the model building. After the inclusion of the eighth term. the change in the value of $R^2$ is found to be only $5*10^{-5}$, which is less than the *threshold value* of 0.001. After the completion of the *forward pass*, both the *generalized* $R^2$ (GRSq) and the $R^2$ converged to a common value of 0.993. During the *backward pass*, the algorithm is not able to prune any term, and all the seven terms used in the forward pass are finally retained in the model. In *Case 1*, the model retains three predictors out of a total of ten predictors. The selected predictors in decreasing order of their importance in the model are found to be: *close_perc*, *high_perc*, and *low_perc*. On the other hand, the predictors which are excluded from the model are: *month*, *day_month*, *day_week*, *time*, *vol_perc*, *nifty_perc*, and *range_diff*. On completion of the execution of the algorithm, the values of some of the important metrics of the model are found to be as follows: (i) *generalized cross-validation* (GCV): 0.0065, *residual sum of square* (RSS): 4.7006, GRSq: 0.9928, and $R^2$: 0.9930. The seven terms that are used by the MARS model under *Case I* are found to be as follows: (i) the *intercept*, (ii) *h*(-0.83682 – *high_perc*), (iii) *h*(*high_perc* – 0.83682), (iv) *h*(-0.692841 – *low_perc*), (v) *h*(*low_perc* – 0.692841), (vi) *h*(-2.11268 – *close_perc*), and (vii) *h*(*close_perc* – 2.11268). For *Case I*, the MARS regression model yields 9 cases out of a total of 745 cases that exhibit a mismatch in sign between the *predicted* and the *actual* values of *open_perc*. The RMSE value for this case is 0.0794, while the mean of the absolute values of the *actual open_perc* is 0.6402. Hence, the ratio of the RMSE to the mean of the absolute values of the *actual open_perc* is 2.4065. The correlation test yields a value of the correlation coefficient as 0.99 with the *t*-statistic value of 325.41, and an associated *p*-value of $2.2*10^{-16}$. This indicates that there is a strong linear relationship between the *predicted* and the *actual* values of *open_perc*. Table 10 presents the results of the MARS regression model. Fig. 13 presents results of MARS regression under *Case I*.

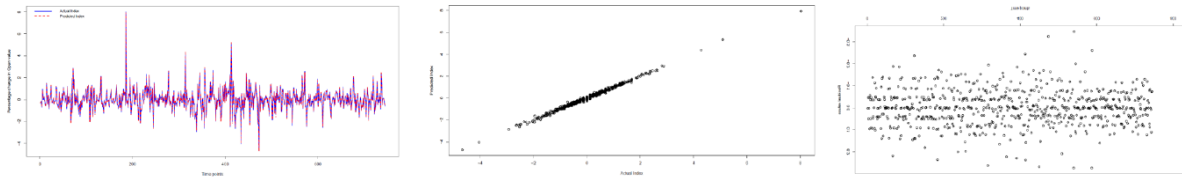

Figure 13: MARS – (a) *actual* and *predicted* values of *open_perc* for *Case I*, (b) relationship between *actual* and *predicted open_perc* for *Case I*, (c) *residuals* for *Case I*

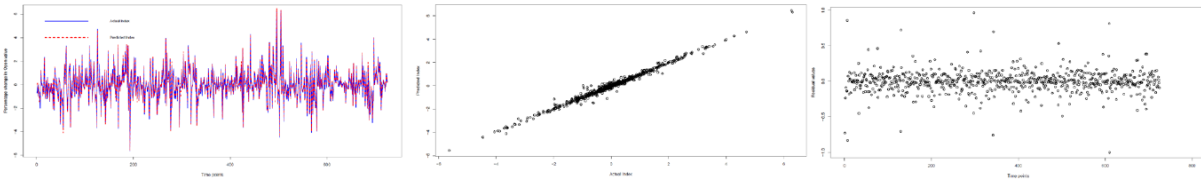

Figure 14: MARS – (a) *actual* and *predicted* values of *open_perc* for *Case II*, (b) the relationship between *actual* and *predicted open_perc* for *Case II*, (c) *residuals* for *Case II*

In *Case II*, the algorithm used nine terms during its *forward execution*. The change in the $R^2$ value at the end of the ninth term is found to be only 0.0002, which is less than the threshold value of 0.001. On completion of the *forward pass*, the values of GRSq and $R^2$ are found to be 0.985 and 0.986 respectively. During the *backward pass* of its execution, the algorithm prunes one term out of the nine terms that were included earlier during the *forward pass*. Hence, the algorithm, finally, uses eight terms in building the regression model. We also observe that the algorithm retains four predictors out of a total of ten predictors that were available initially. The four predictors that are retained in the model, in the decreasing order of their importance, are: *low_perc*, *close_perc*, *range_diff*, and *high_perc*. After the model completes its execution of the *backward pass* the values of some important metrics are noted. They are as follows: GCV: 0.0262, RSS: 18.2512, GRSq: 0.9852, and $R^2$: 0.9858. In Case II, the algorithm uses the following eight terms: (i) the *intercept*, (ii) *h*(0.3675 – *high_perc*), (iii) *h*(*high_perc* – 0.3675), (iv) *h*(-2.6685 – *low_perc*),

(v) $h(low\_perc – 2.6685)$, (vi) $h(0.3996 – close\_perc)$, (vii) $h(-1.8 – range\_diff)$, and (viii) $h(range\_diff - -1.8)$. In *Case II*, we find that 31 cases out of a total of 725 cases exhibit mismatched signs between the *predicted* and the *actual* values of *open_perc*. With an RMSE value of 0.1587, and the mean of the absolute values of the *actual open_perc* as 0.9286, their ratio is found to be 17.09. The correlation test yields the value of the correlation coefficient as 0.99, the *t*-statistic as 223.87, an associated *p*-value of $2.2*10^{-16}$. The high value of the correlation coefficient, and the negligible support for the null hypothesis in the form of a very low *p*-value, indicates that there is a very strong linear relationship between the predicted and the actual values of *open_perc*. Fig. 14 presents results of MARS regression under *Case II*.

The MARS model in *Case III* is identical to that of *Case I*. The model is, however, tested on the 2014 data. We observe that in *Case III*, the model yields 46 cases out of a total of 725 cases that exhibit a sign mismatch between the *predicted* and the *actual open_perc* values. The RMSE of the model, in this case, is found to be 0.1894, while the mean of the absolute values of the actual *open_perc* is 0.9286. The ratio of the RMSE to the mean value is found to be 20.40. The correlation test on the *predicted* and the *actual* values of *open_perc* yields a correlation coefficient value of 0.99, the value of *t*-statistic as 187.13, and an associated *p*-value of $2.2*10^{-16}$. The results indicate that the predicted and the *actual* values of *open_perc* exhibit a strong linear relationship between them for *Case III* as well. The performance results of the MARS model are presented in Fig. 15.

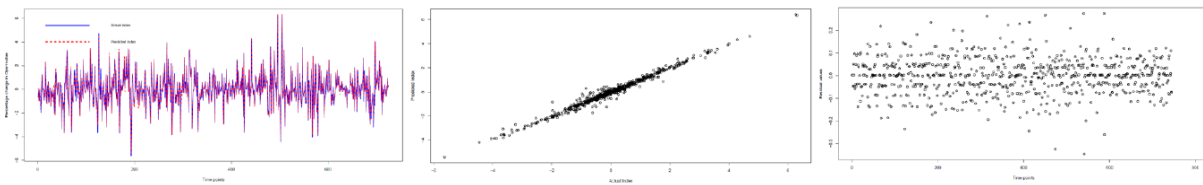

Figure 15: MARS – (a) *actual* and *predicted* values of *open_perc* for *Case III*, (b) the relationship between *actual* and *predicted open_perc* for *Case III*, (c) *residuals* for *Case III*

Table 11: Decision Tree regression results

| Metrics | Case I Training 2013 | Case II Training 2014 | Case III Test 2014 |
|---|---|---|---|
| Correlation Coefficient | 0.97 | 0.97 | 0.10 |
| RMSE/Mean of Absolute Values of Actuals | 35.35 | 37.04 | 165.92 |
| Percentage of Mismatched Cases | 13.42 | 17.38 | 47.72 |

***Decision Tree Regression***: We use the *tree* function defined in the *tree* library in the R programming language to build decision tree-based regression models. For *Case I*, *close_perc* turns out to be the splitting variable at the root node. Other important variables that lead to splitting at nodes are: *high_perc* and *low_perc*. Fig. 16 depicts the decision tree model under *Case I*, and the patterns exhibited by the *actual* and the *predicted* values of the *open_perc* for the model. It is evident from Fig 16(b) that barring some minor deviations, the *predicted* values of *open_perc* very closely follow the pattern of the *actual* values. RMSE for this case is 0.2263, and the mean of the absolute values of the *actual open_perc* is 0.6402. Among the total of 745 cases, 100 cases exhibit a sign mismatch between the *predicted* and the *actual* values of

*open_perc*. The correlation coefficient between the *predicted* and the *actual open_perc* turns out to be 0.97. The *t*-statistics for the correlation test yields a value of 111.35 with a *p*-value of $2.2*10^{-16}$, indicating a strong linear relationship between the *predicted* and the *actual open_perc* values.

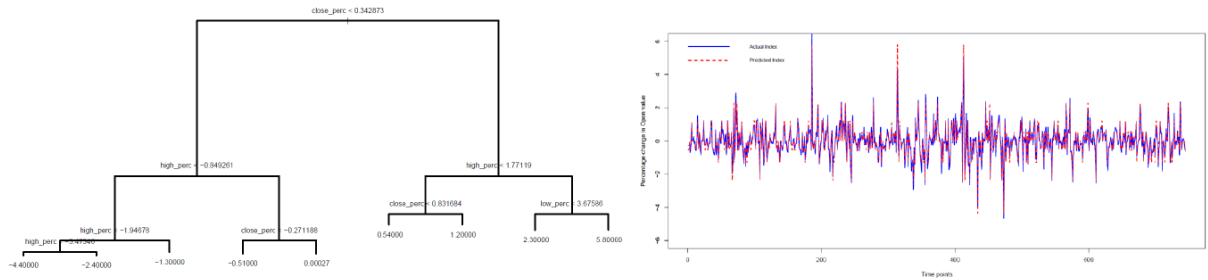

Figure 16: Decision tree regression – (a) the *tree model* for *Case I*, (b) *actual* and *predicted open_perc* for *Case I*

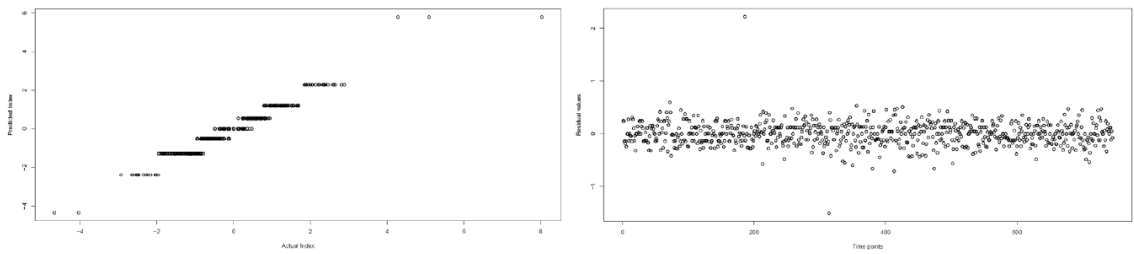

Figure 17: Decision tree regression – (a) the relationship between *actual* and *predicted open_perc* values for *Case I*, (b) *residuals* of regression for *Case I*

Fig. 17 depict some additional performance results of the decision tree-based regression model for *Case I*. Fig 17(a) shows that with the increase in the *actual open_perc* values, its predicted values also exhibit an upward trend, *stepwise*. Fig 17(b) shows that residuals do not exhibit any significant autocorrelation. Table 11 depicts the results of the decision tree regression model.

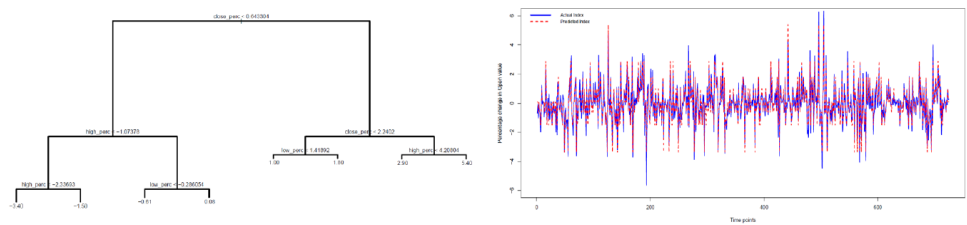

Figure 18: Decision tree regression – (a) the *tree model* for *Case II*, (b) *actual* and *predicted open_perc* for *Case II*

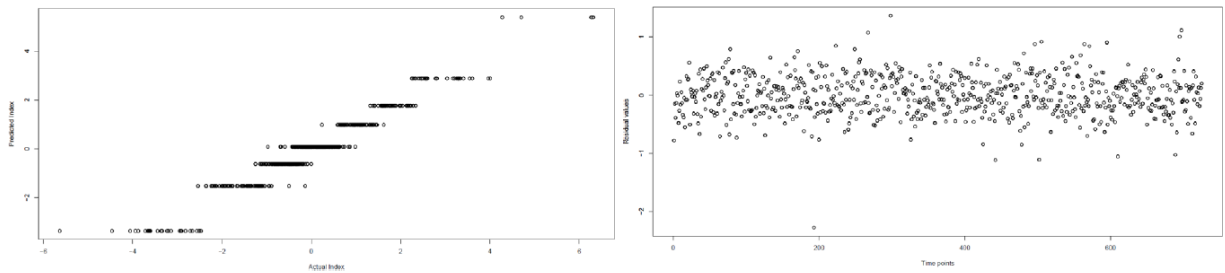

Figure 19: Decision tree regression – (a) the relationship between *actual* and *predicted open_perc* values for *Case II*, (b) *residuals* of regression for *Case II*

Fig. 18 presents the decision tree regression model for *Case II*, and the pattern of variations of the *actual* and the *predicted* values of *open_perc* for this model. The *predicted* values of *open_perc* are found to closely follow the actual values. The variable *close_perc* is again the node that is split at the *root* node. The other two variables which are split subsequently in the decision tree are *high_perc* and *low_perc*. The model for this case yields an RMSE value of 0.3440, while the mean of the absolute values of the *actual open_perc* values is 0.9286. 126 cases out of a total of 725 cases exhibit a sign mismatch between their *predicted* and *actual open_perc* values. The correlation coefficient between the *actual* and the *predicted* values of *open_perc* is found to be 0.96, with the *t*-statistics value of 100.47, and its associated *p*-value of $2.2*10^{-16}$. The *predicted* and the *actual open_perc* values are highly correlated. Fig. 19 shows that the regression model is effective in establishing a linear relationship between the response variable, *open_perc*, and the other predictor variables.

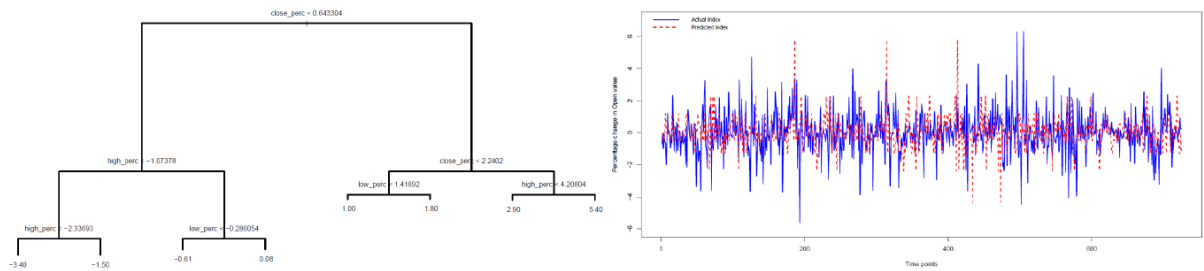

Figure 20: Decision tree regression – (a) the *tree model* for *Case III*, (b) *actual* and *predicted open_perc* for *Case III*

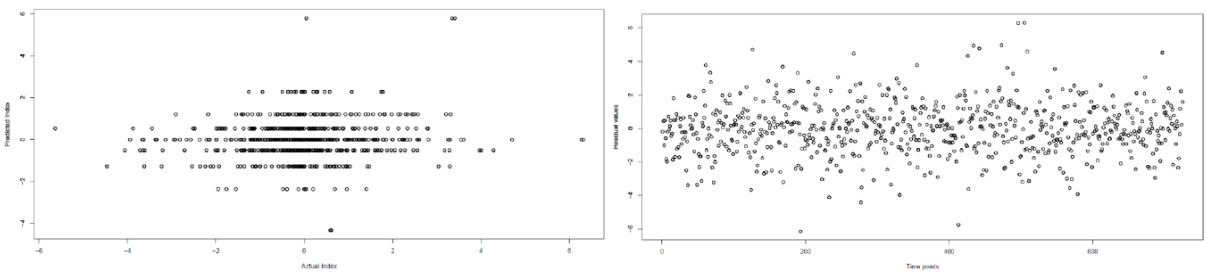

Figure 21: Decision tree regression – (a) the relationship between *actual* and *predicted open_perc* values for *Case III*, (b) *residuals* of regression for *Case III*

The decision tree regression model for *Case III* is identical to that of *Case I* as they are both built on the same dataset. Fig. 20 depicts the decision tree model for *Case III*, and the pattern of variation of the *actual* and *predicted open_perc* values for the model. Even though the model under *Case III* is identical to that under *Case I*, the performance results of the model are different. The model under *Case III* is tested on the 2014 data, while under *Case I*, the model is tested on the 2013 data. The correlation coefficient between the *predicted* and the *actual* values of *open_perc* for this case is found to be 0.10, with the *t*-statistics value of 2.8243, and its associated *p*-value of 0.00487. However, the RMSE for this case is higher than those in the previous two cases. The RMSE is found to be 1.5407, while the mean of the absolute values of the

*actual open_perc* is 0.9286. Thus, the ratio of the RMSE to the mean of the absolute values of *actual open_perc* turns out to be 165.92 – a very high value. 346 cases out of a total of 725 cases exhibit a mismatch in sign between the *predicted* and the *actual* values of *open_perc*. The model is unable to make any accurate prediction as it has a very limited number of values to map into a set of a large set of continuously varying *open_perc* values for the year 2014.

Fig. 21 presents the performance of the model in *Case III*. It is evident from Fig 21 that there are significant deviations between the patterns exhibited by the *actual* and the *predicted open_perc* values. This leads to a significantly higher RMSE in this case for the decision tree model.

Table 12: Bagging regression results

| Metrics | Case I | Case II | Case III |
|---|---|---|---|
|  | Training 2013 | Training 2014 | Test 2014 |
| Correlation Coefficient | 0.96 | 0.98 | 0.97 |
| RMSE/Mean of Absolute Values of Actuals | 40.29 | 25.70 | 34.91 |
| Percentage of Mismatched Cases | 4.70 | 5.10 | 9.24 |

***Bagging Regression*:** The *bagging* function defined in the *ipred* library of R programming language is used in building the bagging regression model. In *Case I*, the RMSE is found to be 0.2579, with the mean of the absolute values of *open_perc* as 0.6402. Out of 745 total cases, 35 cases exhibit mismatch in the *predicted* and the *actual* values of *open_perc*. Fig. 22 presents the performance of the bagging regression model under *Case I*.

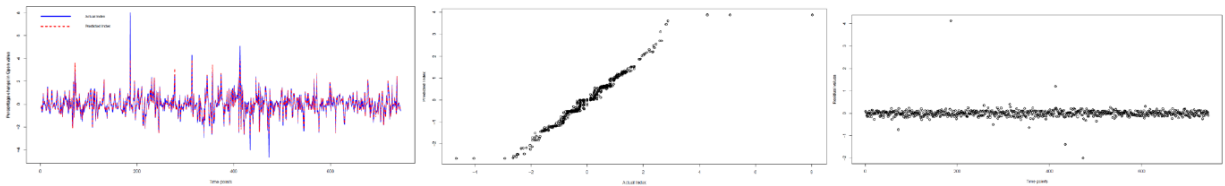

Figure 22: Bagging regression – (a) *actual* and *predicted* values of *open_perc* for *Case I*, (b) relationship between *actual* and *predicted open_perc* for *Case I*, (c) *residuals* for *Case I*

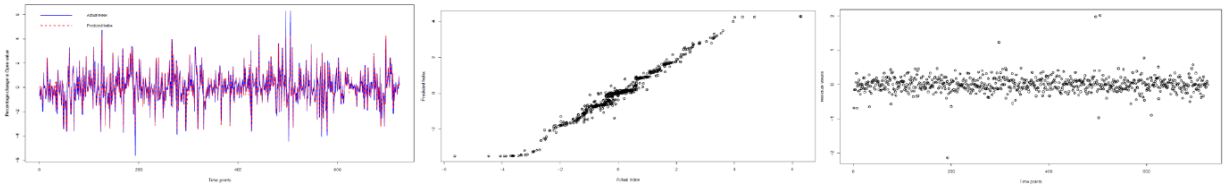

Figure 23: Bagging regression – (a) *actual* and *predicted* values of *open_perc* for *Case II*, (b) the relationship between *actual* and *predicted open_perc* for *Case II*, (c) *residuals* for *Case II*

The bagging model under *Case II* yields an RMSE value of 0.2386, while the mean of the absolute values of the *actual open_perc* is 0.9286. We find that 37 cases out of a total of 725 cases exhibit a mismatch in sign between the *predicted* and the *actual* values of *open_perc*. Fig. 23 depicts the performance of the bagging regression model under *Case II*. The RMSE value for *Case III* is found to be 0.3242. We observe that 67 cases out of a total of 725 cases show a mismatch in sign between its *predicted* and the *actual* values of *open_perc*. The performance of the bagging model under *Case III* is presented in Fig. 24. Table 12 presents the overall results of the bagging regression model.

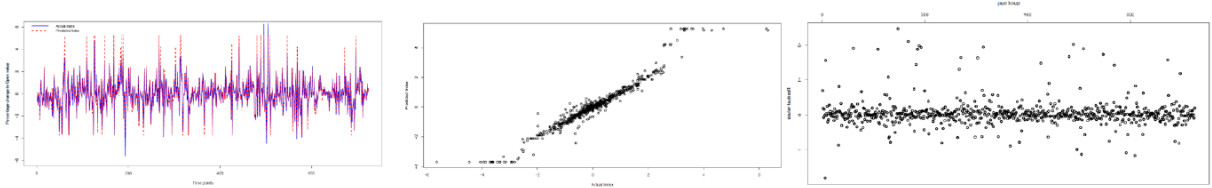

Figure 24: Bagging regression – (a) *actual* and *predicted* values of *open_perc* for *Case III*, (b) the relationship between *actual* and *predicted open_perc* for *Case III*, (c) *residuals* for *Case III*

Table 13: Boosting regression results

| Metrics | Case I<br>Training 2013 | Case II<br>Training 2014 | Case III<br>Test 2014 |
|---|---|---|---|
| Correlation Coefficient | 0.99 | 0.99 | 0.97 |
| RMSE/Mean of Absolute Values of Actuals | 23.40 | 17.35 | 41.51 |
| Percentage of Mismatched Cases | 0.81 | 4.69 | 6.90 |

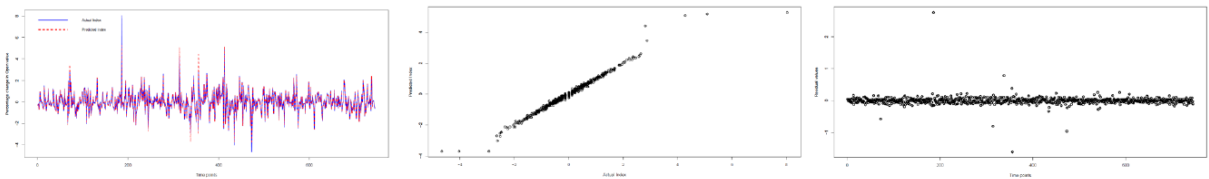

Figure 25: Boosting regression – (a) *actual* and *predicted* values of *open_perc* for *Case I*, (b) relationship between *actual* and *predicted open_perc* for *Case I*, (c) *residuals* for *Case I*

**Boosting Regression**: We use the *blackboost* function defined in the *mboost* library in the R programming language for building the boosting regression models. In *Case I*, 6 cases out of 745 cases exhibit mismatched signs between the *predicted* and the *actual open_perc* values. The RMSE for this case is found to be 0.1498, while the mean of the absolute values of the *actual open_perc* is 0.6402. The performance of the boosting regression model under *Case I* is presented in Fig. 25.

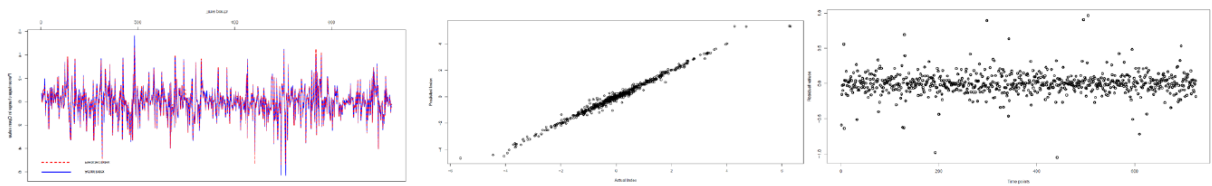

Figure 26: Boosting regression – (a) *actual* and *predicted* values of *open_perc* for *Case II*, (b) the relationship between *actual* and *predicted open_perc* for *Case II*, (c) *residuals* for *Case II*

For *Case II*, out of 725 total cases, 34 cases yield mismatched signs between the *actual* and the *predicted* values of *open_perc*. The RMSE for this case is 0.1611, and the mean of the absolute values of the *actual open_perc* is 0.9286. The performance of the boosting model under *Case II* is presented in Fig. 26.

*Case III* yields an RMSE value of 0.3855, while the mean of the absolute values of the *actual open_perc* is 0.9286. In *Case III*, 50 cases out of a total of 725 cases exhibit mismatched signs between the *predicted* and the *actual open_perc* values. The performance of the boosting model under *Case III* is presented in Fig. 27. Table 13 presents the results of the boosting regression model.

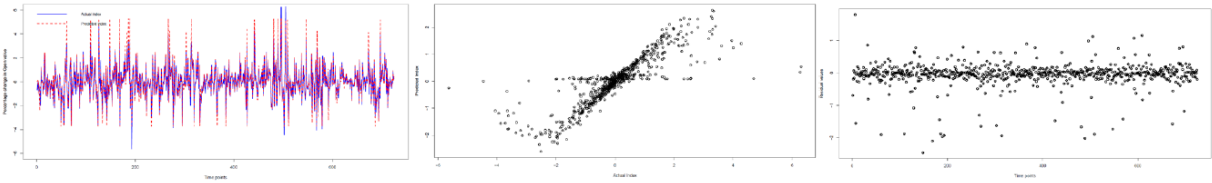

Figure 27: Boosting regression – (a) *actual* and *predicted* values of *open_perc* for *Case III*, (b) the relationship between *actual* and *predicted open_perc* for *Case III*, (c) *residuals* for *Case III*

Table 14: Random Forest regression results

| Metrics | Case I<br>Training 2013 | Case II<br>Training 2014 | Case III<br>Test 2014 |
|---|---|---|---|
| Correlation Coefficient | 0.99 | 0.99 | 0.97 |
| RMSE/Mean of Absolute Values of Actuals | 16.26 | 10.82 | 32.02 |
| Percentage of Mismatched Cases | 0.00 | 2.62 | 6.48 |

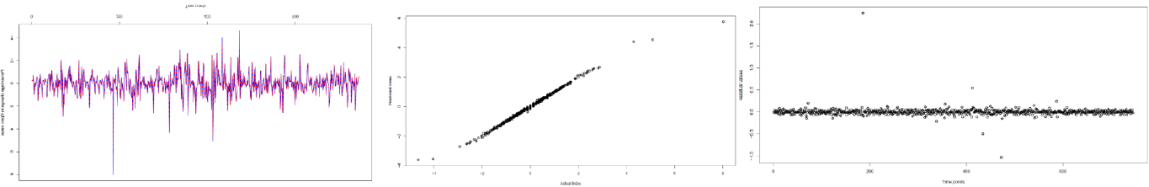

Figure 28: Random forest regression – (a) *actual* and *predicted* values of *open_perc* for *Case I*, (b) relationship between *actual* and *predicted open_perc* for *Case I*, (c) *residuals* for *Case I*

**Random Forest Regression:** We use the *randomForest* function defined in the *randomForest* library in the R programming language for building the random forest regression model. For all three cases, the algorithm tried using three variables at each split of the associated decision tree. The number of regression decision trees constructed in each case is 500. The mean squared residual values are found to be 0.0441, 0.0512, and 0.0441 respectively for *Case I*, *Case II,* and *Case III* respectively. In *Case I*, the percentage of variance explained by the model is 95.13. None of the 745 cases exhibit any sign mismatch between the *predicted* and the *actual* values of *open_perc*. While the RMSE for this case is 0.1041, the mean of the absolute values of the *actual open_perc* is 0.6402. For *Case II*, the model can explain 97.11% of the variance, and

19 cases out of a total of 725 cases exhibit mismatched signs between the *predicted* and the *actual open_perc* values. The RMSE for this case is 0.1005, while the mean of the absolute values of the *actual open_perc* is 0.9286. In *Case III* the model can explain 95.13% of the variance of the response variable. It is observed that 47 cases out of 725 cases exhibit mismatched signs between the *predicted* and the *actual open_perc* values. The RMSE is 0.2973, while the mean of the absolute values of the *actual open_perc* values is 0.9286. Table 14 presents the results of the random forest regression model.

Fig. 28 depicts the performance results of the random forest regression model under *Case I*. Fig. 28(a) depicts the pattern of variation of the *predicted open_perc* values vis-a-vis their corresponding *actual* values for each of the 745 time-slots in *Case I*. The linear relationship between the *predicted* and the *actual open_perc* values is presented in Fig 28(b). The residual values for the regression model are depicted in Fig. 28(c). These three visuals along with the results presented in Table 14 indicate that the random forest regression for *Case I* is very accurate.

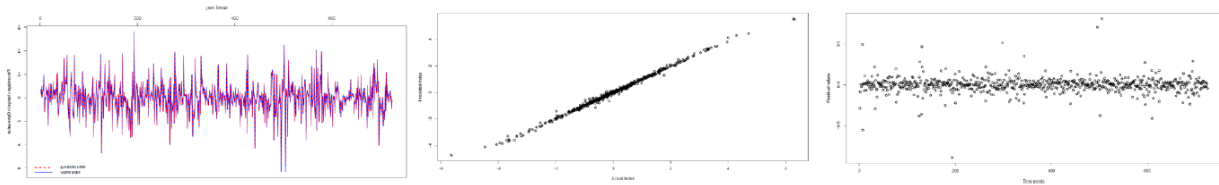

Figure 29: Random forest regression – (a) *actual* and *predicted* values of *open_perc* for *Case II*, (b) relationship between *actual* and *predicted open_perc* for *Case II*, (c) *residuals* for *Case II*

Fig. 29 presents the graphical representations of the performance results of the random forest regression model for *Case II*. It is evident that the *predicted* values of *open_perc* very closely resemble the patterns exhibited by the *actual* values. Moreover, the residuals of the regression model exhibit randomness, and there is no evidence of any autocorrelations in the residuals.

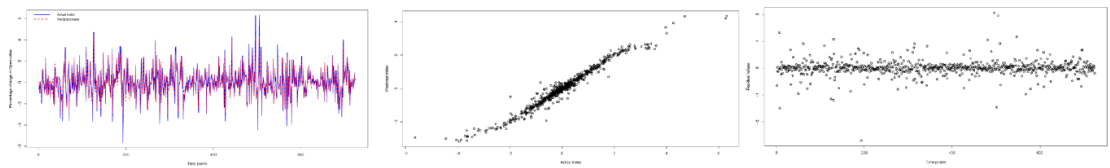

Figure 30: Random forest regression – (a) *actual* and *predicted* values of *open_perc* for *Case III*, (b) relationship between *actual* and *predicted open_perc* for *Case III*, (c) *residuals* for *Case III*

The performance results of the random forest regression model are resented in Fig. 30. The results indicate that the random forest regression model for *Case III* is accurate. Fig 30(b) exhibits some deviations from linearity at the head and the tail portions of the graph, while the middle part of the segment depicts a strong linear pattern between the *actual* and the *predicted* values of *open_perc*. However, the deviations from the linear pattern results in a marginally higher value for the ratio of the RMSE to the mean of the absolute values of *open_perc* under *Case III*.

***ANN Regression*:** We use the *neuralnet* function defined in the *neuralnet* library in the R programming language for designing the ANN regression models. For *Case I*, 9 out of 745 records are found to exhibit mismatched signs between the *actual* and the *predicted open_perc* values. The RMSE, in this case, is found to be 0.1099, and the mean of the absolute values of the *actual open_perc* values is 0.6402. In *Case II*, it is found that 68 out of 725 cases exhibit a mismatch in sign between their *actual* and *predicted open_perc* values. The RMSE of the model for *Case II* is found to be 0.2915. In *Case III*, we find that 79 out of 725 cases exhibit a mismatch in sign between the *predicted* and the *actual* values of *open_perc*. The RMSE of the model for this case is found to be 0.3420. The results for the ANN regression model are presented in Table 15.

Table 15: ANN regression results

| Metrics | Case I<br>Training 2013 | Case II<br>Training 2014 | Case III<br>Test 2014 |
|---|---|---|---|
| Correlation Coefficient | 0.99 | 0.98 | 0.98 |
| RMSE/Mean of Absolute Values of Actuals | 17.16 | 31.39 | 36.83 |
| Percentage of Mismatched Cases | 1.21 | 9.38 | 10.90 |

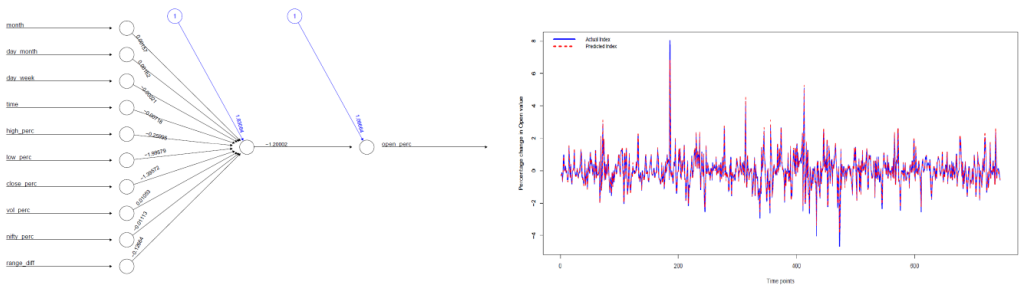

Figure 31: ANN regression – (a) the model for *Case I*, (b) *actual* and *predicted open_perc* for *Case I*

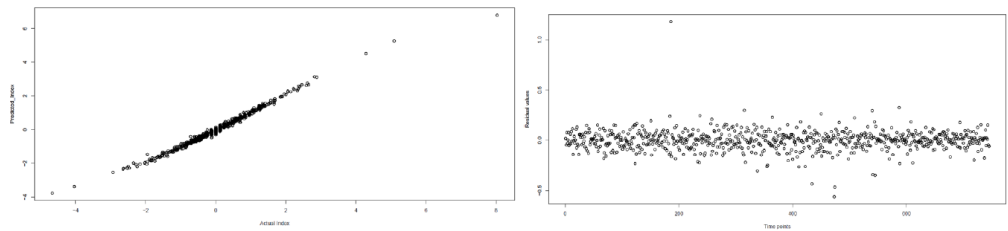

Figure 32: ANN regression – (a) the relationship between *actual* and *predicted open_perc* for *Case I*, (b) *residuals* for *Case I*

Fig. 31 depicts the ANN regression model for *Case I*, and the pattern of variation of *actual* and *predicted* values of *open_perc*. Only one node is used in the *hidden layer* as it is found to be sufficient to model the nonlinearity in the data. In the figure, the link weights are marked in *black* color, while the bias values associated with the hidden layer node and the output layer node are written in *blue* color. The nodes in the input layer correspond to the input variables of the model. Fig. 32 exhibits the functional relationship between the *predicted* and the *actual* values of *open_perc*, and the residuals of the regression model. It is evident that the residuals are random, and they do not exhibit any significant autocorrelation. For *Case I*, the correlation coefficient between the *actual* and the *predicted* values of *open_perc* is 0.99, and the percentage of cases that exhibit sign mismatch between the *predicted* and the *actual open_perc* is 1.07.

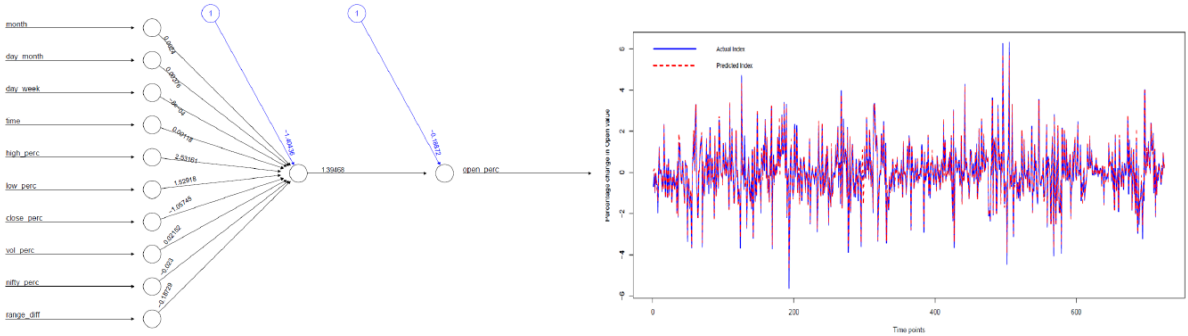

Figure 33: ANN regression – (a) the model for *Case II*, (b) *actual* and *predicted open_perc* for *Case II*

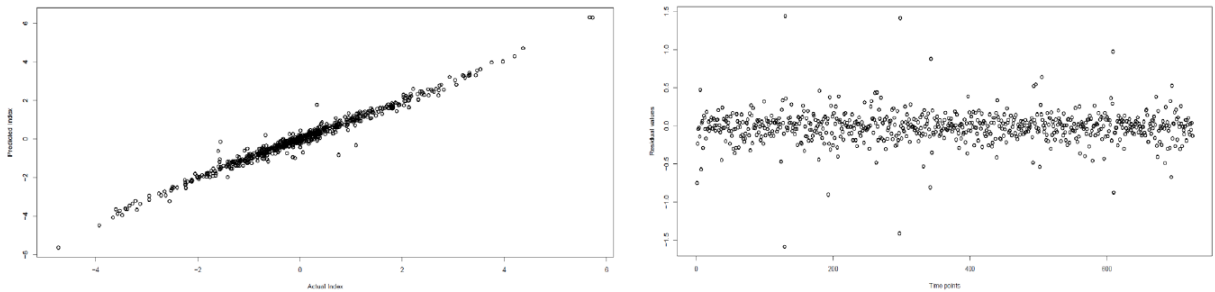

Figure 34: ANN regression – (a) the relationship between *actual* and *predicted open_perc* for *Case II*, (b) *residuals* for *Case II*

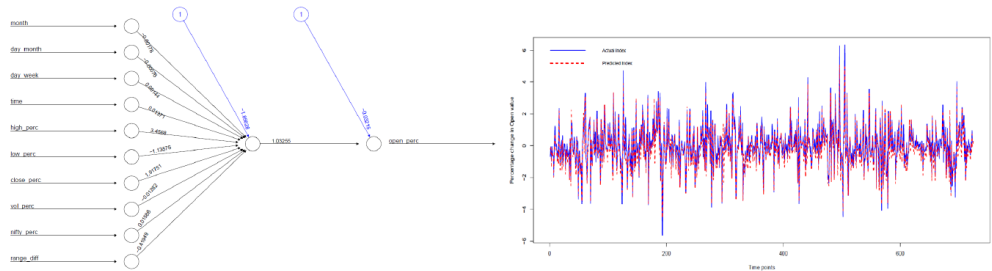

Figure 35: ANN regression – (a) the model for *Case III*, (b) *actual* and *predicted open_perc* for *Case III*

Fig. 33 shows the ANN regression model for *Case II,* and the pattern of variation of the *actual* and the *predicted* values of *open_perc* for this model. It is evident that the *predicted* values of *open_perc* very closely follow the *actual open_perc* values. Fig. 34 depicts that there is a *high degree of linearity* in the relationship between the *predicted* and the *actual* values of *open_perc*, and the residuals of the regression model are free from any significant autocorrelation.

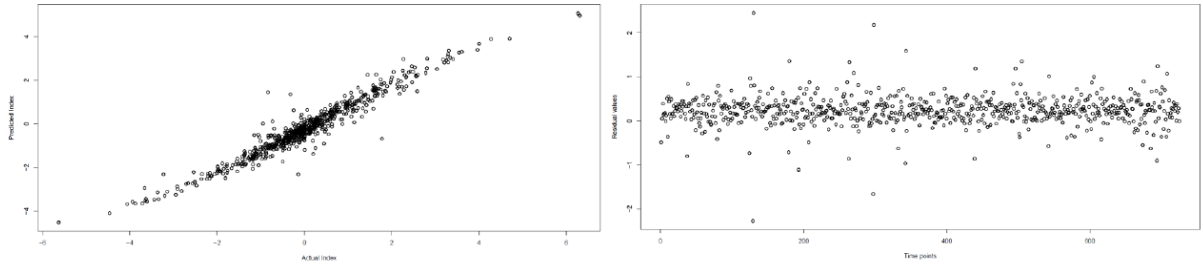

Figure 36: ANN regression – (a) the relationship between *actual* and *predicted open_perc* for *Case III*, (b) *residuals* for *Case III*

Fig, 35 and Fig. 36 depict the ANN regression model for *Case III* and its various performance results. These figures and the values of the numerical metrics like *correlation coefficient*, the ratio of RMSE to the mean of the absolute values of the *actual open_perc*, and the number sign mismatch cases between the *predicted* and the *actual* values of *open_perc*, all show that the model is very accurate in its task of prediction.

Table 16: SVM regression results

| Metrics | Case I Training 2013 | Case II Training 2014 | Case III Test 2014 |
| --- | --- | --- | --- |
| Correlation Coefficient | 0.93 | 0.98 | 0.83 |
| RMSE/Mean of Absolute Values of Actuals | 53.88 | 27.92 | 82.96 |
| Percentage of Mismatched Cases | 0.27 | 4.41 | 13.19 |

***SVM Regression***: We use the *svm* function defined in the *e1071* library of the R programming language. For all three cases, the regression type used is *eps-regression*, and the *kernel* is *radial*. The values of the parameters *gamma* and *epsilon* are both found to be 0.1. The algorithm finds the number of *support vectors* as 248, 265, and 246 for *Case I*, *Case II*, and *Case III* respectively. The RMSE values for the three cases are found to be 0.3450, 0.2593, and 0.7703 respectively. The mean of the absolute values of the *open_perc* is 0.6402. We compute the ratio of the RMSE to the mean of the absolute values of *open_perc* for all three cases. We also identify the cases which exhibit a difference in the signs between the *actual* and the *predicted* values of *open_perc*. These are the cases, in which the regression model fails to predict the direction of the movement of the *actual open_perc* values.

For *Case I*, 2 out of 745 cases are found to exhibit a sign *mismatch* between the *actual* and the *predicted* values of *open_perc*. 32 out of 725 cases yield sign mismatch in *Case II*. In *Case III*, since the model encounters a more difficult challenge in prediction, 95 out of 725 cases are mismatched in sign. The SVM regression results are presented in Table 16. For all three cases, the SVM regression model looks quite accurate in its results.

Fig. 37 presents the performance results of the SVM regression model under *Case I*. It is evident from Fig. 37(a) that for this model, the *predicted* series very closely follows the pattern of the variation in the *actual open_perc* values. Fig. 37 (b) clearly shows that except for some points at the tail and the head, most of the points exhibit a strong linear relationship between the *actual* and the *predicted* values of *open_perc* for *Case I*. The residual plot in Fig. 37(c) shows that most of the residuals values are random.

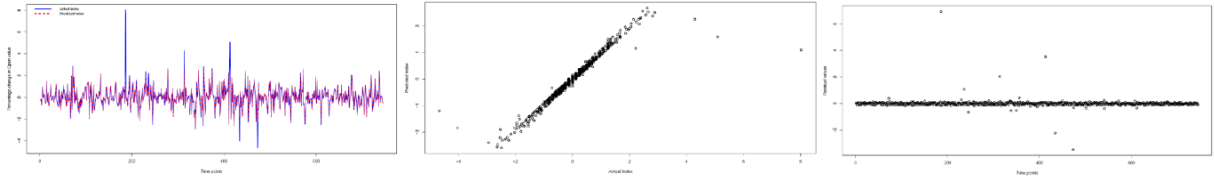

Figure 37: SVM regression – (a) *actual* and *predicted* values of *open_perc* for *Case I*, (b) relationship between *actual* and *predicted open_perc* for *Case I*, (c) *residuals* for *Case I*

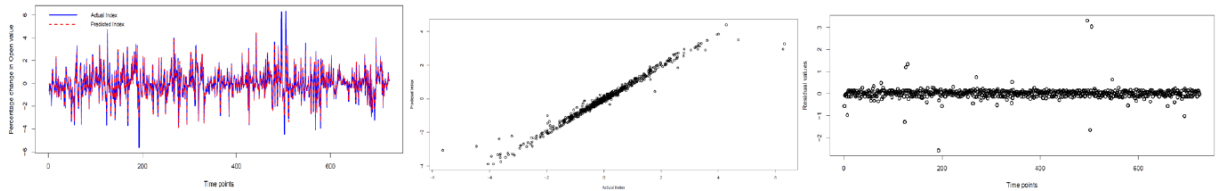

Figure 38: SVM regression – (a) *actual* and *predicted* values of *open_perc* for *Case II*, (b) the relationship between *actual* and *predicted open_perc* for *Case II*, (c) *residuals* for *Case II*

The performance of the SVM model in *Case II* is very similar to that in *Case I*. This is evident from Fig. 38. These figures exhibit similar patterns shown in Fig. 37. However, it is found that the ratio of the RMSE to the mean of the absolute values of the *actual open_perc* is smaller in *Case II* than that in *Case I*.

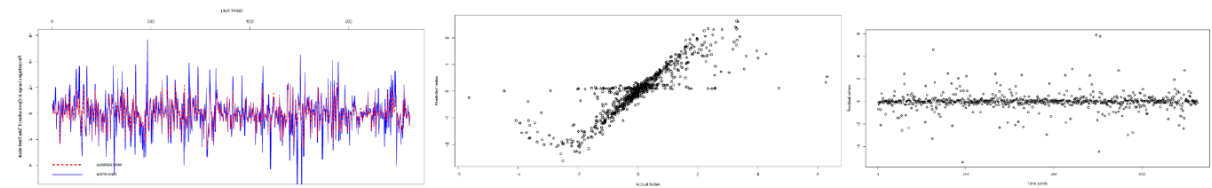

Figure 39: SVM regression – (a) *actual* and *predicted* values of *open_perc* for *Case III*, (b) the relationship between *actual* and *predicted open_perc* for *Case II*, (c) *residuals* for *Case III*

However, it is observed from Fig. 39 that *Case III* poses a more difficult challenge to the SVM regression model. For *Case III*, the correlation coefficient between the *actual* and the *predicted* values of the *open_perc* is found to be much smaller in comparison to the other two cases. Fig. 39(a) shows that the *predicted* values of *open_perc* do not exactly follow the pattern of variation exhibited by the *actual open_perc* values. It is also evident from Fig. 39(b) that there is a high degree of nonlinearity between the *predicted* and the *actual open_perc* values. However, no evidence of any significant autocorrelations between the residuals are observed in Fig. 39(c).

***LSTM Regression***: In Section 5, we briefly presented a brief overview of LSTM networks. In the following, we present, the results of the LSTM-based regression models. These are the sequential steps that are followed in building the LSTM models: (i) *reading* the file containing the raw data, (ii) *normalizing* the data, (iii) *converting* the normalized data into a *time series* and then into a *supervised learning* problem, (v) *creating* a *deep learning model* using Tensorflow and Keras frameworks, (vi) *training* and *validating* the model, (vii) *visualization* of the training and validation performance, and (viii) *evaluating* the *accuracy* of the model on test data. The raw data consists of the following attributes: (i) *year*, (ii) *month*, (iii) *day*, (iv) *hour* (i.e., the *time slot*), (v) *open*, (vi) *high*, (vii) *low*, (viii) *close*, (ix) *volume*, and (x) the NIFTY *index*. Using *Python* programming, we combine the attributes *year*, *month*, *day*, and *hour* into a single attribute, and use the combined new attribute as the *index*. Other remaining attributes are now used for building the models. We provide the details of the three cases in the following.

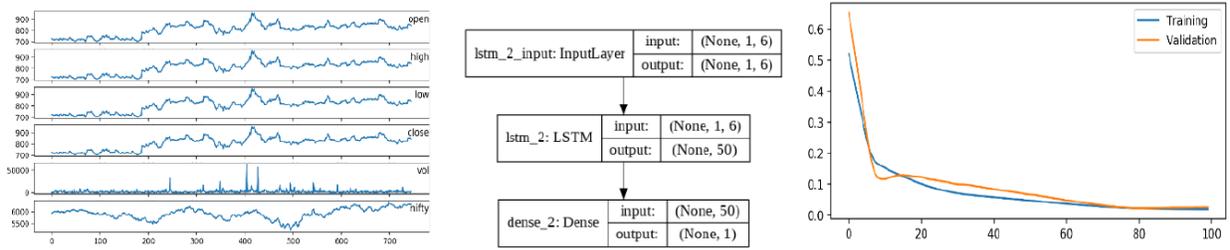

Figure 40: LSTM regression – (a) the representation of the *stock price data* for *Case I*, (b) the *model architecture* for *Case I*, *Case II*, and *Case III*, and (c) the *training* and *validation losses* for *Case I*

For *Case I*, we first plot the *open*, *high*, *low*, *close*, *volume,* and the NIFTY time series. In this case, there are 746 records. Fig. 40(a) depicts the time series for each of the attributes. All the six *value* attributes (leaving out the *index* column) are then *normalized* using the *MinMaxScalar* function defined in the *processing* submodule of the *sklearn* module in *Python*. Out of the total 746 records, the first 500 records are used for training the model while the remaining 246 records are left out for validation. The *Sequential* function defined in *Keras* is used for building the layers and the model is compiled using *mean absolute error* (MAE) as the *loss function* and *adaptive moment estimation* (Adam) as the optimizer. The model architecture is presented in Fig. 40(b). In this model, the input layer consists of six attributes, and the input data is passed on to the LSTM layer that has 50 nodes so that 50 features are extracted from the input data. The output of the LSTM layer is passed on to a *dense layer* (i.e., a *fully connected layer*) that has 1 node at its output which produces the next predicted value of the time series. The behavior of the training and the validation loss values is studied for different values of *epochs* and *batch sizes*. With a batch size of 72 and an epoch value of 100, the training and validation losses are found to have converged to a very low value. Fig. 40(c) presents the behavioral patterns of the training and the validation losses in *Case I*. On the completion of the final epoch, the RMSE value is 8.812, and the *correlation coefficient* between the *actual* and the *predicted open* values is 0.983. The *training* and *validation loss* values are found to be 0.0194 and 0.0252 respectively.

*Case II* involves *slot-wise* stock prices for the year 2014, and there are 725 records in the dataset in this case. Fig. 41(a) presents the plots for the time series attributes for *Case II*. In this case too, before building the model, the raw values of the six attributes are *normalized* using the *MinMaxScalar* function. The model architecture remains identical to that of *Case I*, which is represented in Fig. 40(b). For training the model, the first 500 records are used, while the remaining 225 records are used for the model validation. It is observed that, initially, the validation loss converges with the training loss at an epoch value of 40.

However, it starts increasing again with the increase in the epoch value. The validation loss finally converges with the training loss at an epoch value of 100 when a batch of 72is used. The RMSE of the model is found to be 15.002, and the correlation coefficient between the *actual* and the *predicted open* values is 0.982. On completion of the final epoch, the training and the validation loss are found to be 0.0134 and 0.0301 respectively. Fig. 41(b) depicts the convergence of the training and the validation loss in *Case II*.

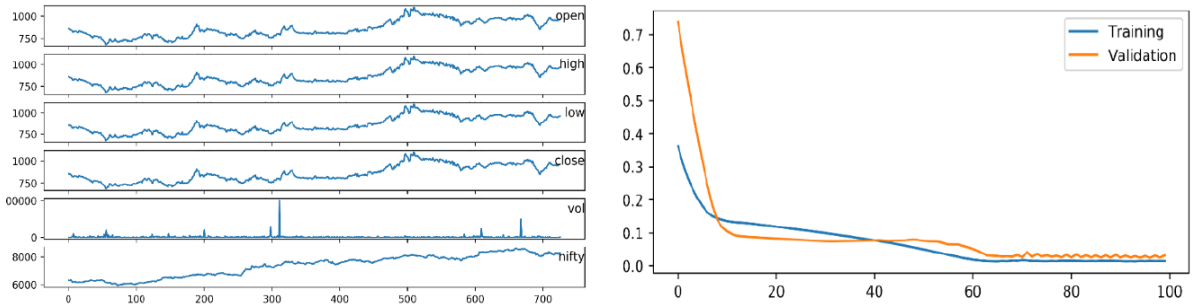

figure 41: LSTM regression – (a) the representation of the *stock price data* for *Case II*, (b) the *training* and *validation losses* for *Case II*

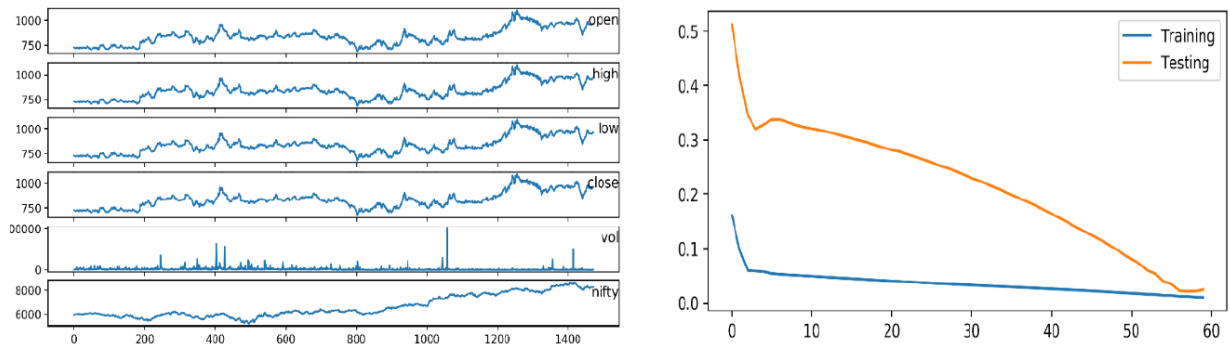

Figure 42: LSTM regression – (a) the representation of the *stock price data* for *Case III*, (b) the *training* and *validation losses* for *Case III*

In *Case III*, the LSTM model is built using data of the year 2013, and then the model is tested on the stock price records of the year 2014. The raw dataset, in this case, consists of 1471 records in total - 746 records (those belonging to the year 2013) are used in building the model, and the remaining 725 records (those belonging to the year 2014) are used for testing the model. Fig 42(a) presents the plots of the *open*, *high*, *low*, *close*, *volume*, and the NIFTY time series for the 1471 records. The LSTM model architecture remains unchanged and is identical to that of *Case I*. The training and the test losses converge with each other after an epoch value of 60 with a batch size of 72. The RMSE and the correlation coefficient, for this case, are found to be 13.477 and 0.996 respectively. After the completion of the last epoch, the training and the test losses are 0.0116 and 0.0258 respectively. Fig, 42(b) depicts the convergence of the training and the testing losses with varying epoch values for *Case III*.

***CNN Regression*:** In Section 5, we presented a brief overview of how CNN regression models are used to carry out multi-step forecasting of the open values of the stock price time series. Unlike the machine learning and the LSTM models, we follow a different approach in model building testing for the CNN models. We use the stock price data from December 31, 2012 (a Monday) to January 9, 2015 (a Friday) for our analysis. The original stock price dataset consists of records that are collected at 5 minutes interval of time. For each time slot the values of *open*, *high*, *low*, *close*, and *volume* are available. The stock price data from December 31, 2012, to December 30, 2013, is used for training, while the models are tested on the data from December 31, 2013 to January 9, 2015. The time-series data is also arranged in the form of a repetitive weekly sequence from Monday to Friday. Fig. 43 depicts the time series data of stock prices at 5 minutes. As mentioned in Section 5, we follow four different models of CNN regression. The models are as follows.

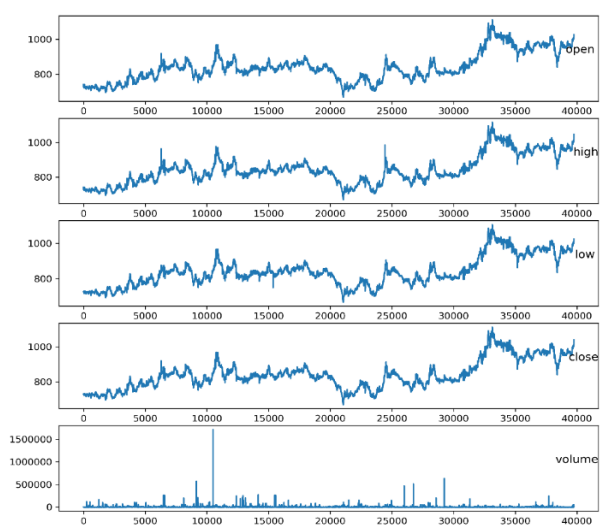

Figure 43: The stock price data of Godrej Consumer Ltd. at five minutes interval during 2013 - 2015

***CNN Model #1*:** This model is based on a *univariate forecasting approach with the previous one week's data* as the input. Since is a week consists of five data points, the input size of the model is five. Since the input data size is small, we build a light CNN model for this case. One *convolution layer* with 16 filters and a kernel size of 3 is used. In other words, it means that the input sequence of five days is read with a convolutional operation in three time-steps at a time, and this operation is performed 16 times. A *max-pooling layer* of size 2 is used that reduces the size of the feature maps before the internal representation is *flattened* to one long vector. This is interpreted by a *fully-connected* layer before the output layer (which is also fully connected) predicts the *open* values for the next five days. Fig. 44(a) depicts the architecture of the CNN model #1. In the convolution layer and the fully connected layer, the *rectified linear unit* (ReLU) function is used as the *activation function*. The *Adam* optimizer is used in the stochastic gradient descent algorithm, and 20 epochs with a batch size of 4 are used for building and compiling the model. The *mean squared error* (MSE) is used as the loss function at the output layer. For computing the error in prediction, *root mean squared error* (RMSE) is used as the metric. With a small batch size, and with the use of the stochastic nature of the gradient descent algorithm, the model is expected to learn a slightly different mapping of the inputs to the outputs every time it is trained. This implies that performance results will vary slightly in different runs. Hence, the model is iterated over several rounds and the mean values of the metrics are taken as the overall performance of the model.

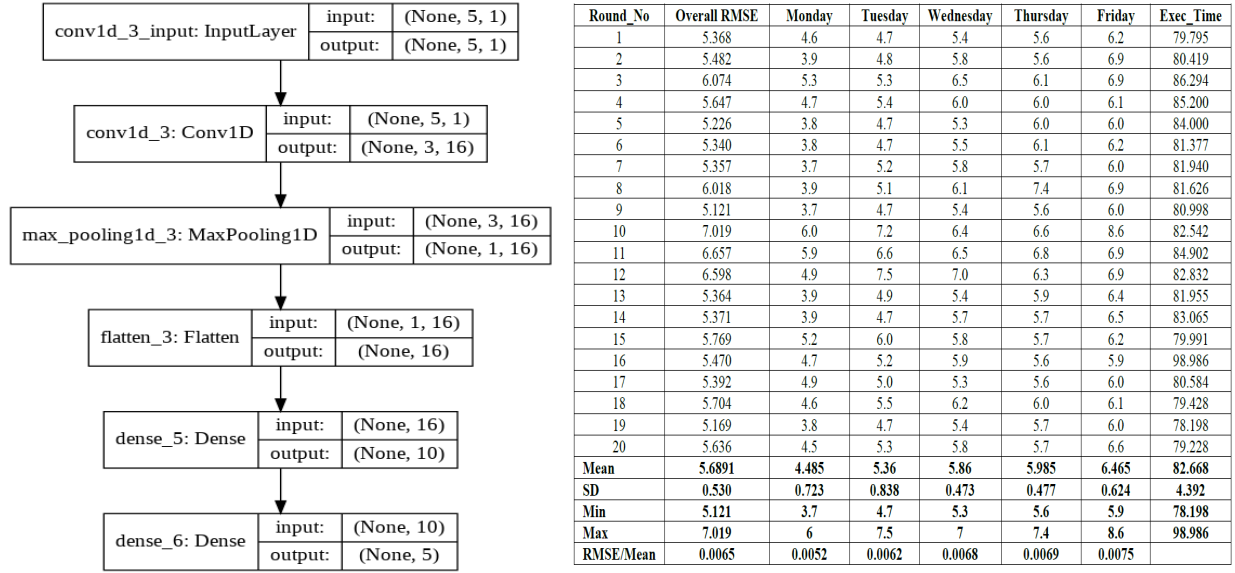

Figure 44: The CNN Model #1 – (a) the *model architecture*, and (b) regression results of the *univariate time series model with previous one week's data as the input*

We test the model for 20 rounds and note the performance of the model in each round for its overall RMSE, the RMSE values for the individual days of a week (i.e., Monday – Friday), the execution time of the model, and the ratio of the RMSE to the mean value of the variable predicted (i.e., mean of the *open* value for the test dataset). Figure 44(b) presents the results for the performance of the CNN model for *Case I*. The mean value of *open* for the test data is 866.5875. The training and the test data consist of 19500 and 20250 records respectively. The execution time for the model is expressed in *seconds*. The model is executed on a system consisting of an Intel i7 CPU with a clock speed of 2.60 GHz- 2.59 GHz and 16 GB *random access memory* (RAM) running on the Windows 10 operating system.

***CNN Model #II:*** This model is based on univariate forecasting with the previous two week's data as the input. The architecture of this model is identical to that of *CNN Model #I*. However, this model is fed with two weeks' prior data (i.e., 10 immediate past *open* values) to forecast the *open* values of the next week. Fig. 45(a) depicts the architecture of the CNN model for *Case II*. The performance results of the CNN Model #II for 20 rounds of its execution are presented in Fig. 45(b).

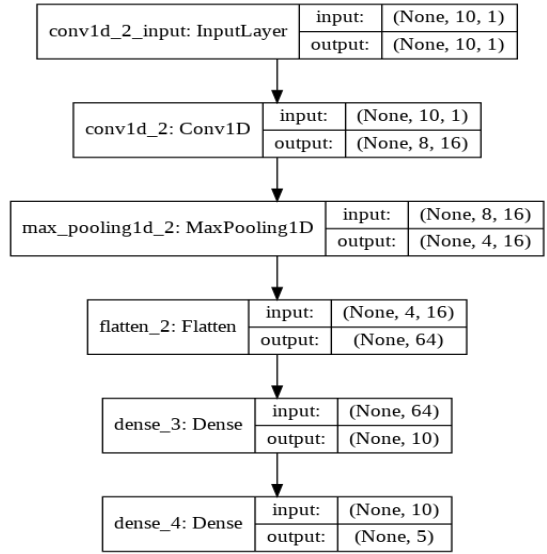

| Round_No | Overall RMSE | Monday | Tuesday | Wednesday | Thursday | Friday | Exec_Time |
|---|---|---|---|---|---|---|---|
| 1 | 5.307 | 3.9 | 4.6 | 5.6 | 5.7 | 6.4 | 85.386 |
| 2 | 6.042 | 3.9 | 6.2 | 6.4 | 6.8 | 6.6 | 85.845 |
| 3 | 5.378 | 4.0 | 5.0 | 5.6 | 5.6 | 6.3 | 84.247 |
| 4 | 5.278 | 3.5 | 4.6 | 5.5 | 5.9 | 6.3 | 108.214 |
| 5 | 5.592 | 4.4 | 5.0 | 5.7 | 6.0 | 6.7 | 88.994 |
| 6 | 8.852 | 6.6 | 8.7 | 9.6 | 10.4 | 8.5 | 83.484 |
| 7 | 5.294 | 3.8 | 5.3 | 5.5 | 5.7 | 6.0 | 85.899 |
| 8 | 6.061 | 4.7 | 4.9 | 5.9 | 6.8 | 7.5 | 88.373 |
| 9 | 5.229 | 3.9 | 4.9 | 5.4 | 5.8 | 5.8 | 88.297 |
| 10 | 5.857 | 5.6 | 5.3 | 5.3 | 5.9 | 7.1 | 84.636 |
| 11 | 5.227 | 4.8 | 4.7 | 5.6 | 5.4 | 5.7 | 101.964 |
| 12 | 8.797 | 7.8 | 8.4 | 9.0 | 9.5 | 9.1 | 87.828 |
| 13 | 7.190 | 5.3 | 6.6 | 7.9 | 8.5 | 7.2 | 81.649 |
| 14 | 5.697 | 4.9 | 5.7 | 5.8 | 5.9 | 6.1 | 84.072 |
| 15 | 5.314 | 3.6 | 4.5 | 5.6 | 6.3 | 6.1 | 87.396 |
| 16 | 5.186 | 3.6 | 4.8 | 5.3 | 5.7 | 6.2 | 81.858 |
| 17 | 7.110 | 9.3 | 6.0 | 6.5 | 7.1 | 6.2 | 88.401 |
| 18 | 5.356 | 3.7 | 4.5 | 5.2 | 6.0 | 6.8 | 84.992 |
| 19 | 5.210 | 4.1 | 4.6 | 5.3 | 5.6 | 6.2 | 84.526 |
| 20 | 6.889 | 6.7 | 6.6 | 7.1 | 6.9 | 7.1 | 82.084 |
| Mean | 6.043 | 4.905 | 5.545 | 6.190 | 6.575 | 6.695 | 87.407 |
| SD | 1.145 | 1.578 | 1.228 | 1.261 | 1.370 | 0.871 | 6.528 |
| Min | 5.186 | 3.500 | 4.500 | 5.200 | 5.400 | 5.700 | 81.649 |
| Max | 8.852 | 9.300 | 8.700 | 9.600 | 10.400 | 9.100 | 108.214 |
| RMSE/Mean | 0.0070 | 0.0057 | 0.0064 | 0.0071 | 0.0076 | 0.0077 | |

Figure 45: The CNN Model #2 – (a) the *model architecture*, and (b) regression results of the *univariate time series model with the previous two weeks' data as the input*

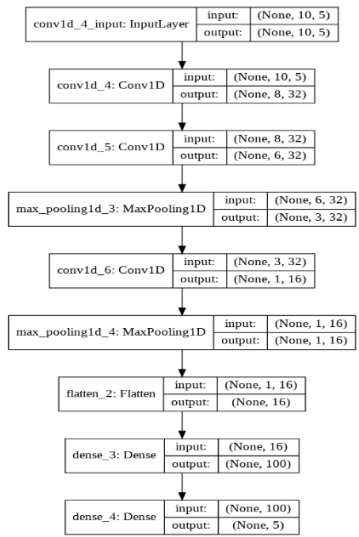

| Round_No | Overall RMSE | Monday | Tuesday | Wednesday | Thursday | Friday | Exec_Time |
|---|---|---|---|---|---|---|---|
| 1 | 8.466 | 7.9 | 6.7 | 7.4 | 9.6 | 10.1 | 134.286 |
| 2 | 5.510 | 4.2 | 5.2 | 5.6 | 6.1 | 6.3 | 122.020 |
| 3 | 6.003 | 4.7 | 5.6 | 6.3 | 6.4 | 6.8 | 111.729 |
| 4 | 6.718 | 5.6 | 6.4 | 7.0 | 7.2 | 7.3 | 116.582 |
| 5 | 5.602 | 4.4 | 5.2 | 5.7 | 6.1 | 6.4 | 131.704 |
| 6 | 6.056 | 4.3 | 5.9 | 6.2 | 6.7 | 6.8 | 113.107 |
| 7 | 5.585 | 4.3 | 5.0 | 5.7 | 5.9 | 6.8 | 137.406 |
| 8 | 6.220 | 4.6 | 6.6 | 6.5 | 6.4 | 6.8 | 113.899 |
| 9 | 5.710 | 4.5 | 5.4 | 5.9 | 6.1 | 6.4 | 113.717 |
| 10 | 6.101 | 5.1 | 6.0 | 6.2 | 6.3 | 6.7 | 111.018 |
| 11 | 6.708 | 6.7 | 6.6 | 6.5 | 6.6 | 7.1 | 131.253 |
| 12 | 5.398 | 4.1 | 5.0 | 5.5 | 5.9 | 6.2 | 139.208 |
| 13 | 7.956 | 8.3 | 6.4 | 6.4 | 7.8 | 10.2 | 114.197 |
| 14 | 7.061 | 6.5 | 6.3 | 6.4 | 8.4 | 7.5 | 115.814 |
| 15 | 5.870 | 4.5 | 5.6 | 6.1 | 6.4 | 6.5 | 116.391 |
| 16 | 6.070 | 4.7 | 5.7 | 6.2 | 6.5 | 7.0 | 113.899 |
| 17 | 5.977 | 5.2 | 5.4 | 6.0 | 6.4 | 6.7 | 116.089 |
| 18 | 5.760 | 4.8 | 5.0 | 6.5 | 6.1 | 6.2 | 114.035 |
| 19 | 5.647 | 4.9 | 5.2 | 5.7 | 6.0 | 6.3 | 112.558 |
| 20 | 5.547 | 4.4 | 5.1 | 5.6 | 5.9 | 6.5 | 116.758 |
| Mean | 6.198 | 5.185 | 5.715 | 6.17 | 6.64 | 7.03 | 119.784 |
| SD | 0.819 | 1.218 | 0.601 | 0.488 | 0.949 | 1.124 | 9.307 |
| Min | 5.398 | 4.1 | 5 | 5.5 | 5.9 | 6.2 | 111.018 |
| Max | 8.466 | 8.3 | 6.7 | 7.4 | 9.6 | 10.2 | 139.208 |
| RMSE/Mean | 0.0072 | 0.0060 | 0.0066 | 0.0071 | 0.0077 | 0.0081 | |

Figure 46: The CNN Model #3 – (a) the *model architecture*, and (b) regression results of the *multivariate time series model with the previous two weeks' data as the input*

**CNN Model #III**: This model is based on a *multivariate forecasting approach with the previous two week's data* as the input. In this case, the multi-channel CNN model uses all the values of the variables – *open*, *high*, *low*, *close*, and *volume* – over the previous two weeks for forecasting the next week's *open* values. This is done by allowing each variable of the time series to enter into the model as a separate channel of input. In this case, CNN uses a separate kernel, and reads each input sequence into a separate set of features, and learning from those features. The increase in the amount of data requires a larger and more sophisticated model that is trained for a longer time. Two convolutional layers, each with 32 filter maps and a kernel size of 3 are used. The two convolutional layers are followed by a *max-pooling layer* of size 2. The output of the *max-pooling layer* is passed onto a third *convolutional layer* with 16 filter maps with a kernel size of 3. The third convolutional layer passes its output to a *max-pooling layer of size* 1. A fully connected layer with 100 nodes then interprets the features before finally sending the predicted values to the five nodes in the output layer. The model is trained over 70 epochs with a batch size of 16. The *activation function* for all layers is ReLU, and the *Adam* optimizer is used for optimizing the *stochastic gradient descent* algorithm. Fig. 46(a) depicts the architecture of the CNN Model #3. The performance results of the model are presented in Fig. 46(b).

From Figs. 44, 45, and 46, it is evident observed that the CNN Model #1 – the univariate CNN model with the previous one week's data as the input is the most accurate. It yields a ratio of RMSE to the mean of the *actual* values as 0.0065. This model is also found to be the fastest in execution with a mean execution time of 82.668 seconds for 20 rounds. The CNN Model #III – the multivariate CNN model with the previous two weeks' data as the input, is found to have performed the worst. This model yields a value of 0.0072 for the ratio of the RMSE to the mean of the *actual* values. The model also takes the longest time to execute. The mean execution time per round for this model is found to be 119.784 seconds for the 20 rounds over which the model is trained. It is also noted that for all the models, the ratio of RMSE to the mean value of the response variable is the lowest for Monday, and the same is the highest for Friday. The ratio of RMSE to the mean *open* value consistently increased from Monday through Friday.

We now present a summary of the performance results for all the machine learning-based classification and regression results.

***Overall Performance***: Finally, we summarize the performance of the machine learning and LSTM models different predictive models. Tables 21 – 23 present the performance of the classification models for *Case I*, *Case II*, and *Case III* respectively. For each case and each metric, the best model is marked with a bold font.

Table 21: Summary of the performance of the classification models in *Case I*

|  | LR | KNN | DT | BAG | BOOST | RF | ANN | SVM |
|---|---|---|---|---|---|---|---|---|
| Sensitivity | 94.79 | 89.57 | 95.09 | 95.09 | **100.00** | 94.48 | 95.40 | 94.46 |
| Specificity | 97.61 | 96.42 | 98.09 | 98.09 | **100.00** | 97.61 | 97.61 | 95.58 |
| PPV | 96.87 | 95.11 | 97.48 | 97.48 | **100.00** | 96.86 | 96.88 | 94.17 |
| NPV | 96.01 | 92.24 | 96.25 | 96.25 | **100.00** | 98.08 | 96.46 | 98.09 |
| CA | 96.38 | 93.42 | 96.78 | 96.78 | **100.00** | 96.24 | 96.64 | 96.38 |
| F1 Score | 95.82 | 92.26 | 96.27 | 96.07 | **100.00** | 95.66 | 96.13 | 94.31 |

Table 22: Summary of the performance of the classification models in *Case II*

|  | LR | KNN | DT | BAG | BOOST | RF | ANN | SVM |
|---|---|---|---|---|---|---|---|---|
| Sensitivity | 94.83 | 86.93 | 92.40 | 95.44 | **100.00** | 93.01 | 93.62 | 94.67 |
| Specificity | 95.96 | 92.93 | 95.71 | 96.46 | **100.00** | 94.19 | 95.71 | 93.35 |
| PPV | 95.12 | 91.08 | 94.70 | 95.73 | **100.00** | 93.01 | 94.77 | 91.79 |
| NPV | 95.72 | 89.54 | 93.81 | 96.22 | **100.00** | 94.19 | 94.75 | 95.71 |
| CA | 95.45 | 90.21 | 94.21 | 96.00 | **100.00** | 93.66 | 94.76 | 93.93 |
| F1 Score | 94.97 | 88.96 | 93.54 | 95.58 | **100.00** | 93.01 | 94.19 | 93.21 |

We observe that both for *Case I* and *Case II* and all the metrics, *boosting* performed the best among all the classification models. However, because *Case I* and *Case II* represent only the training scenarios, the performance in *Case III* that depicts the performance in the testing phase should be considered as the most critical. From Table 23, it is observed that the ANN model performs the best on *sensitivity* and NPV, while the *boosting* model outperforms all other models on *specificity*, PPV, and *classification accuracy*. However, the random forest model is found to have performed best on the *F1 score* – the metric that is usually considered to be the most important one in classification. In Tables 21-26, the following abbreviations are used in the column names: LR – *logistic regression*, KNN – *k-nearest neighbor*, DT- *decision tree*, BAG – *bagging*, BOOST – *boosting*, RF – *random forest*, ANN – *artificial neural networks*, SVM – *support vector machines*, MV – *multivariate regression*, MARS – *multivariate adaptive regression spline*, LSTM – *long- and- short- term memory*.

Table 23: Summary of the performance of the classification models in *Case III*

|  | LR | KNN | DT | BAG | BOOST | RF | ANN | SVM |
|---|---|---|---|---|---|---|---|---|
| Sensitivity | 92.10 | 84.50 | 89.97 | 89.97 | 92.10 | 91.19 | **99.70** | 93.81 |
| Specificity | 89.39 | 48.99 | 92.42 | 92.42 | **93.43** | 92.93 | 34.60 | 90.19 |
| PPV | 87.83 | 57.92 | 90.80 | 90.78 | **92.10** | 91.46 | 55.88 | 87.54 |
| NPV | 93.16 | 79.18 | 91.73 | 91.73 | 93.43 | 92.70 | **99.28** | 95.20 |
| CA | 90.62 | 65.10 | 91.31 | 91.31 | **92.83** | 92.14 | 64.14 | 91.72 |
| F1 Score | 89.91 | 68.73 | 90.38 | 90.37 | 92.10 | **91.32** | 71.62 | 90.57 |

Table 24: Summary of the performance of the regression models in *Case I*

|  | MV | MARS | DT | BAG | BOOST | RF | ANN | SVM | LSTM |
|---|---|---|---|---|---|---|---|---|---|
| Correlation | **0.99** | **0.99** | 0.97 | 0.96 | **0.99** | 0.99 | 0.99 | 0.93 | **1.00** |
| RMSE/Mean | 13.32 | **12.41** | 35.35 | 40.29 | 23.40 | 16.26 | 17.16 | 53.88 | **7.94** |
| Mismatched Cases | 18.67 | 1.21 | 13.42 | 4.70 | 0.81 | **0.00** | 1.21 | **0.27** | **0.00** |

Table 25: Summary of the performance of the regression models in *Case II*

|  | MV | MARS | DT | BAG | BOOST | RF | ANN | SVM | LSTM |
|---|---|---|---|---|---|---|---|---|---|
| Correlation | **0.99** | **0.99** | 0.97 | 0.98 | **0.99** | 0.99 | 0.98 | 0.98 | **1.00** |
| RMSE/Mean | 18.84 | 17.09 | 37.04 | 25.70 | 17.35 | **10.82** | 31.39 | 27.92 | **4.04** |
| Mismatched Cases | 5.38 | 4.28 | 17.38 | 5.10 | 4.69 | **2.62** | 9.38 | 4.41 | **0.00** |

Table 26: Summary of the performance of the regression models in *Case III*

|  | MV | MARS | DT | BAG | BOOST | RF | ANN | SVM | LSTM |
|---|---|---|---|---|---|---|---|---|---|
| Correlation | **0.99** | **0.99** | 0.10 | 0.97 | 0.97 | 0.97 | 0.98 | 0.83 | **0.99** |
| RMSE/Mean | **18.88** | 20.40 | 165.92 | 34.91 | 41.51 | 32.02 | 36.83 | 82.96 | **2.36** |
| Mismatched Cases | **5.24** | 6.34 | 47.72 | 9.24 | 6.90 | 6.48 | 10.90 | 13.19 | **2.40** |

Tables 24 – 26 present the performance results of the regression models, including the LSTM-based deep learning model. Since the LSTM model outperforms all the machine learning models on all the metrics and for all three cases, we also identify the best performing machine learning model on each metric.

In *Case I*, *multivariate regression*, MARS, *boosting*, *random forest*, and ANN all yield the highest correlation coefficient value of 0.99. However, the correlation coefficient is found to be 1.00 in the case of LSTM. For the ratio of the RMSE to the mean of the absolute values of the *open_perc* values, MARS yields the lowest value of 12.41 among the machine learning models, while the corresponding value for LSTM is 7.94. Both random forest and LSTM yield no *sign mismatch* between the *predicted* and the *actual* values of *open_perc*.

In *Case II*, the highest value of the correlation coefficient is achieved by *multivariate regression*, MARS, *boosting*, and *random forest*. LSTM outperforms all the machine learning models on this metric by attaining a value of 1.00. The RMSE to the mean ratio value of 10.82 is the least for random forest among the machine learning models. However, the corresponding value yielded by LSTM was 4.04. Random forest produces only 2.62 percent cases that mismatch in the signs of the *actual* and the *predicted open_perc* values, however for LSTM, all the cases have the same sign for the *actual* and the *predicted open_perc* values.

For *Case III*, while LSTM exhibits the best performance on all metrics, *multivariate regression* and MARS yield the same (the highest) value for the correlation coefficient. For the metric RMSE to the mean ratio, and the percentage of the mismatched cases, *multivariate regression* produces the best results among the machine learning models.

It may be noted that the CNN models are built and tested on the stock price data from December 31, 2012, to January 9, 2015, and the data are of 5 minutes' granularity. The predictions are made on one week's time horizon daily. Since all other models built in this work are based on stock price data aggregated into three slots in a day, it is not wise to compare the performance of the CNN model suites with the machine learning-based models and the LSTM models. However, one can easily see that based on the ratio of the RMSE to the mean of the actual values of the forecasted variable, all the CNN models outperform the LSTM by a large margin. While the least value for the ratio of the RMSE to the actual value of the forecasted variable for the LSTM model was found to be 2.36, the corresponding value for the CNN suite was 0.0065.

## 7. Conclusion and Future Work

In this paper, we propose a robust forecasting framework for stock price and stock price movement patterns with a very high level of accuracy. The predictive model consists of eight classification and eight regression models based on several machine learning approaches. Also, the predictive framework includes two deep learning models of regression built on an LSTM network and a suite of CNNs. The machine learning-based models and the LSTM model use stock price data of three slots in a day and predict the stock price slot-wise over a prediction horizon of one year. We construct the models, train, validate, and finally test them using the historical stock prices of a company – Godrej Consumer Products Ltd. The data is extracted from the listed values of the stock in the National Stock Exchange (NSE) of India during the period of two years – January 2013 till December 2014. The stock price data are extracted at five minutes intervals of time using the Metastock tool. The raw data is pre-processed, appropriate transformation (i.e., normalization, standardization, NA removal, etc.) done, and several derived predictor variables are created based on the rich features of the stock data. While several newly derived predictors are used in building the model, we use the percentage change in the *open* values of the stock, called *open_perc*, as the response variable. The five minutes interval granular data are also aggregated into three slots on a given day so that the predictive models can be built to forecast the value of the *open_perc* in the next slot given stock price data till the current slot. While the classification-based models are used to predict the movement pattern of *open_perc* values, the objective of the regression models is to accurately predict the value of the *open_perc*. In addition to exploiting the machine learning algorithms for building the eight classification and eight regression models, we also leverage the rich features of Tensorflow and Keras frameworks in building two extremely powerful deep learning-based regression models using an LSTM network and a suite of CNNs. For building the machine learning models, we use the R programming language, while for the LSTM-based deep learning regression model, and the suite of four CNN models, the Python programming has been used. The results elicit a very interesting observation. While no single machine learning model performs the best on all the metrics on classification and regression, the deep learning model using an LSTM network outperforms all the machine learning-based regression models on every metric. Since the CNN models are built using stock price data collected at 5 minutes interval of time while the machine learning models and the LSTM models are based on stock price data collected at three slots in a day, it is not appropriate to compare the performance of the CNN suite with the other models. However, it has been found that based on the metric of the ratio of the RMSE to the mean of the actual values of the forecasted variable, the CNN models are far more accurate than the machine learning models and the LSTM-based deep learning models.